\documentclass[twocolumn]{aastex701}

\usepackage{savesym}
\savesymbol{tablenum}

\usepackage{graphicx}%
\usepackage{multirow}%
\usepackage{amsmath,amssymb,amsfonts}%
\usepackage{amsthm}%
\usepackage{mathrsfs}%
\usepackage[title]{appendix}%
\usepackage{xcolor}%
\usepackage{tabularx}

\usepackage{chemformula}
\usepackage{hyperref}
\usepackage[separate-uncertainty=true]{siunitx}
\usepackage{float}
\usepackage{lipsum}
\usepackage{cleveref}

\usepackage{tikz}
\usetikzlibrary{arrows.meta,positioning,fit,calc}

\usepackage[ruled]{algorithm2e}
\usepackage{algorithmic}

\DeclareSIUnit\bar{bar}
\DeclareSIUnit\AU{AU}
\DeclareSIUnit\dex{dex}
\DeclareSIUnit\erg{erg}
\DeclareSIUnit\day{day}
\DeclareSIUnit\year{yr}

\newcommand{\IW}{\text{IW}}
\newcommand{\ppmw}{\text{ppmw}}
\newcommand{\Hppmw}{\text{H}_\ppmw}

\begin{document}

\title{Constraining the lives and times of exoplanets through evolutionary Bayesian retrievals}


\author[orcid=0000-0002-8368-4641]{Harrison Nicholls}
\email[show]{harrison.nicholls@ast.cam.ac.uk}
\affiliation{Institute of Astronomy, University of Cambridge, United Kingdom}

\author[orcid=0000-0002-3286-7683]{Tim Lichtenberg}
\email{tim.lichtenberg@rug.nl}
\affiliation{Kapteyn Astronomical Institute, University of Groningen, The Netherlands}

\author[orcid=0009-0003-9173-9539]{Ben Riegler}
\email{ben.riegler@helmholtz-munich.de}
\affiliation{Helmholtz AI, Technical University of Munich, Munich, Germany}

\author[orcid=0009-0002-9247-2437]{Robb Calder}
\email{rdc49@cam.ac.uk}
\affiliation{Institute of Astronomy, University of Cambridge, United Kingdom}

\author[orcid=0000-0002-0640-2671]{Vincent Fortuin}
\email{probml@utn.de}
\affiliation{Helmholtz AI, Technical University of Munich, Munich, Germany}
\affiliation{University of Technology Nuremberg, Germany}

\begin{abstract}

\noindent Static retrieval frameworks are leading tools for interpreting exoplanet observations, yet time-independent modelling leaves them prone to degeneracy and unable to resolve exoplanets' histories. The compositions and structures of surveyed super-Earth and sub-Neptune sub-populations remain unclear, but are shaped by physics acting across Gyr timescales. Interpreting these planets as static non-evolving snapshots allows multiple degenerate scenarios to explain their observed properties. We develop a generalised parameter retrieval framework, built on asynchronous Bayesian optimisation to efficiently dispatch a multi-physics forward-model, resolving exoplanets' evolving properties from their initial magma ocean conditions to the present day. By building Bayesian retrievals into the PROTEUS framework, sensitive coupled interior-atmosphere interactions are naturally resolved and interpretations are constrained to physically permissible scenarios. We test evolutionary retrievals with three exoplanet prototypes: a young sub-Neptune, an older super-Earth, and a warm terrestrial planet --- representative of the surveyed exoplanet population. Evolutionary retrieval jointly infers their mantle redox conditions, metallic core fractions, and early volatile inventories from spectroscopically accessible observables. Some scenarios remain subject to well-established degeneracies between core fractions and volatile budgets. Terrestrial-mass exoplanets benefit from strong observable-parameter correlations that lift these degeneracies; we recover post-formation volatile inventories with $\lesssim 20$ percent error. Exoplanet science is primed for incoming JWST, PLATO, Roman, and ELT data --- observations which necessitate careful interpretation. Adoption of time-evolved models lifts interpretive degeneracies, providing the means to understand the deep interiors and lifetime histories of worlds throughout our galaxy.
\end{abstract}



\section{Introduction} 
\label{sec:intro}
Exoplanet observations measure present-day atmospheric compositions and bulk properties, such as masses and radii. Insight into specific planet's earlier conditions must be drawn by pairing fingerprints of physical processes imprinted onto their current conditions (e.g., C/O ratios; \cite{bond_the_2010, madhusudhan_co_2012}) against a conceptual understanding of planetary formation and evolution processes \citep{bergin_carbon_2026, krijt_chemical_2023}. In parallel, we can attempt to explain the observed trends in the wider exoplanet population, such as the `radius valley' dividing super-Earths from sub-Neptunes \citep{Fulton_RadVal_2017, ho_deep_2023}, by consideration of the driving physical processes. Computer models necessarily provide the connection between observations and planets' past and internal conditions \citep{barstow_outstanding_2020, madhusudhan_exoplanetary_2019}.

A common approach for \textit{inferring} the conditions within exoplanet interiors, from observations, is to perform a parameter \textit{retrieval} \citep{madhusudhan_retrieval_2009}. Parameter retrievals generally use statistical methods to carefully sample a pre-defined parameter space; attempting to converge upon a set of parameters which best fit some observations. A forward model is used to map parameters to observables. The outcome from evaluating each forward model --- i.e., each `best guess' on a planet's true state --- is a necessary quantity for targeting subsequent iterations of a retrieval process. 

The efficiency of statistical retrieval frameworks has enabled exploration of diverse physical regimes and inference of small exoplanets' internal structures \citep{piaulet_evidence_2023, benneke_toi270d_2024, valencia_diversity_2025}.  However, existing frameworks require computationally fast forward models: a weakness mitigated by the adoption of simplified physics and substantial modelling assumptions \citep{foreman_emcee_2013}. For example, `free chemistry' retrievals permit parametrised atmospheric chemical abundances to be derived from an observation, without specific considerations of chemical favourability  \citep{fisher_HowDoWe_2022, barstow_outstanding_2020, tsai_comparative_2021}. Forward models invoked within retrieval frameworks have largely simulated exoplanets' \textit{present} day structures and compositions \citep{macdonald_catalog_2023}. They do not consider planets' lifetime histories, relevant physical hysteresis effects \citep{turbet_venus_2021, way_venus_2016, boer_absence_2025}, nor whether the modelled conditions are physically achievable \citep{nicholls_beyond_2026, macdonald_why_2020}. For example, it remains unclear whether massive \ch{H2O} envelopes are permissible formation outcomes \citep{kimura_water_2026}. Deep insight from current methods is inhibited by multiple degenerate static-structure solutions being readily able to explain the observed states of individual planets \citep{barstow_outstanding_2020}. For example, sub-Neptunes' bulk densities can be equally explained by steam atmospheres and small metal-core fractions, or extended hydrogen-dominated atmospheres overlying denser interiors \citep{rogers_road_2025, Venturini2020, luque_density_2022, bitsch_water_2019}. 

Computer simulations of planetary formation and evolution provide a physically-justified basis for predicting how planets change throughout their lives \citep{chili_protocol_2025, tonks_magma_1993}. We expect that all planets form in hot, molten states, with large inventories of volatiles obtained from their formation and accretion \citep{norris_volatile_2017, abe_early_1986, kimura_formation_2020}. Young planets cool and lose some fraction of their initial volatiles \citep{owen_review_2019}, subject to regulation by atmospheric greenhouse effects and buffering by interior-atmosphere volatile exchange \citep{nicholls_redox_2024, hamano_emergence_2013, lichtenberg_vertically_2021}.  These planetary evolution models are constructed for \textit{predictive} applications. For example, to demarcate magma ocean versus potentially habitable regimes \citep{krissansen_erosion_2024, wogan_rapid_2022, schaefer_magma_2016}.
Critically, since planetary evolution models are constrained by our knowledge of physics and reasonable assumptions about planets' initial conditions \citep{elkins_linked_2008, schaefer_review_2018}, they may naturally disfavour inferences of unphysical scenarios. Modelled physics and chemistry provide bounds on realistic planetary structures and compositions  \citep{lichtenberg_constraining_2025, madhusudhan_exoplanetary_2016}.

Evolution models have also been applied \textit{interpretively} through simple grid search approaches, which construct grids of models for \textit{post-hoc} comparison against observations \citep{nicholls_escape_2025, krissansen-totton_implications_2023, schaefer_magma_2016}. Grid searches are inefficient approaches for parameter inference, because the grid size scales geometrically with the number of parameters.  Exploratory applications of Bayesian machine learning techniques to parameter inference have demonstrated a factor-of-eight performance improvement over classical retrieval algorithms \citep{pagliaro_exoformer_2026, garvin_machine_2024, hayes_optimizing_2020,yakubu_machine_2026}.  Previously, \citet{owen_bayesian_2016} applied the MESA planetary evolution model within a Bayesian inference framework to estimate the initial atmosphere masses for exoplanets Kepler-36\,b~and~c \citep{paxton_mesa_2011}, suggesting that generalised incorporation of planetary evolution models into retrieval frameworks may be feasible.

Here, we test the efficacy of incorporating physically-comprehensive planetary evolution simulations into a generalised parameter retrieval framework, for inferring exoplanets' historical and interior conditions. The PROTEUS planetary evolution simulation framework is adopted and expanded with a new mode of operation. Previously, PROTEUS could run simulations of planetary evolution either standalone or across regularly sampled parameter-grid axes. Here, we implement a new run mode in PROTEUS to efficiently dispatch simulations (forward models) for parameter retrieval, through asynchronous Bayesian optimisation. We test parameter retrieval with three synthetic exoplanet prototypes, as a potential successor to the contemporary non-evolving static retrieval paradigm: a young sub-Neptune, an older super-Earth, and a warm terrestrial planet. 

This paper is structured as follows:
\begin{itemize}
    \item Section~\ref{sec:methods} describes our framework for simulating planetary evolution, PROTEUS, and develops our asynchronous Bayesian optimisation (ABO) retrieval method. We also describe the static-structure model used for comparison. 

    \item  Section~\ref{sec:results_truth} establishes three ground-truth scenarios, to generate synthetic observables from chosen parameters. We test retrieving these parameters with ABO (Section~\ref{sec:results_bo}) and compare our new approach against static-structure retrievals (Section~\ref{sec:results_static}). We also assess which optimisation algorithms are better suited for this new application.

    \item Section~\ref{sec:discuss_bo} discusses the efficacy of evolutionary retrievals to infer the atmospheric, interior, and historical conditions of different exoplanet classes. Section~\ref{sec:discuss_gaps} considers directions for future development. Section~\ref{sec:discuss_real} contextualises our findings to specific exoplanets and upcoming telescopic surveys. 
    
    \item Section~\ref{sec:conclude} draws conclusions and highlights directions for future research.
\end{itemize}

\section{Methods} 
\label{sec:methods}

\subsection{Planetary evolution modelling}
\label{sec:methods_proteus}

We use the PROTEUS~framework\footnote{\url{https://proteus-framework.org/}} for simulating the time-evolution of planets and their stars. PROTEUS is an established modelling tool for simulating the thermal and compositional evolution of low-mass rocky bodies \citep{nicholls_redox_2024, lichtenberg_vertically_2021}. The framework's modular architecture includes rigorous software testing, and enables PROTEUS' flexible application from sub-Earth to mini-Neptune scenarios  \citep{lichtenberg_proteus_2026, barker_introducing_2022}. PROTEUS self-consistently tracks the time-evolution of a planet's metallic core, mantle, and atmosphere, alongside the radiation emitted by its host star. The atmosphere regulates planet cooling, while the mantle geodynamics track internal energy production, surface temperature evolution, and magma ocean solidification from the core-mantle-boundary upwards \citep{abe_thermal_1993, elkins_linked_2008, bower_linking_2019}.

The mantle begins in an initially molten state --- presumed to have arisen from planetary formation and early giant impact events \citep{tonks_magma_1993, warren_moon_1985, hirschmann_magma_2012}. Energy and mass are redistributed between the atmospheric and interior domains; the interior domain includes a silicate mantle and metallic core. At each time-step, after evolving the mantle geodynamics by sub-stepping through some time interval $\Delta t$, the upper energy-flux boundary condition on the mantle is updated to achieve a self-consistent solution with the simulated atmospheric climate state \citep{nicholls_redox_2024, lichtenberg_vertically_2021, lebrun_thermal_2013}. 

PROTEUS adopts an iterative approach with adaptive time-stepping. Simulations begin with an initial time-step $\Delta t$ of 200\,years. $\Delta t$ is then adapted depending on the rates at which the mantle melt fraction $\Phi_m$ and the outgoing energy flux $F_\mathrm{atm}$ change. The time-step is capped by absolute and relative limits: $\Delta t\le 5\mathrm{\,Myr} + \frac{6}{10}t$. The wall-clock runtime of each PROTEUS simulation is variable, but generally correlates with the physical integration time of the simulation, since the modelled physics imposes a ceiling on the simulation time-step size $\Delta t$. 

In this work, PROTEUS simulations terminate under four independently sufficient criteria: (a) when the requested integration time $t_\mathrm{evo}$ has been attained, or (b) when the atmosphere surface pressure is reduced to $<1\mathrm{\,bar}$ through escape processes, or (c) when the mantle melt fraction $\Phi_m$ is less than 5\,wt\% \citep{nicholls_redox_2024}, or (d) when the simulation wall-clock runtime exceeds 600\,seconds (Appendix~\ref{app:evaltime}). The relevant input and output variables to these PROTEUS calculations are described in Section~\ref{sec:methods_params}.

Atmospheric compositions are re-calculated at each iteration by PROTEUS' outgassing module, CALLIOPE\footnote{\url{https://proteus-framework.org/CALLIOPE/}}, using an equilibrium thermochemical-solubility scheme \citep{sossi_redox_2020, bower_retention_2022}. CALLIOPE solves for partial pressures of the major volatile species\footnote{\ch{CO2}, \ch{CO}, \ch{H2O}, \ch{N2}, \ch{SO2}, \ch{S2}, \ch{H2S}, \ch{H2}, \ch{NH3}, \ch{O2}} subject to H-C-N-S mass conservation between the planet's atmosphere and mantle \citep{sossi_solubility_2023, nicholls_redox_2024}. These volatile species are partitioned between the molten mantle reservoir and outgassed atmosphere subject to the empirically derived solubility laws, which depend on the temperature and mantle redox conditions \citep{sossi_solubility_2023, boulliung_so2_2022, dixon_co2_1997, armstrong_co_2015, ardia_ch4_2013, dasgupta_nitrogen_2022, gaillard_redox_2022}. These volatiles' dissolved components are assumed to be controlled by magma saturation at the surface, so they become progressively degassed as the planet cools and the available reservoir of molten mantle material decreases \citep{nicholls_chili_2026, schaefer_redox_2017, katyal_effect_2020, walbecq_outgassing_2025}. Simultaneously, we solve for volatile species' partial pressures at thermochemical equilibrium using temperature-dependent reaction coefficients derived from JANAF \citep{JANAF}.

The chemical redox conditions of the upper-mantle factor into our volatile partitioning and chemistry calculations. Mantle redox is largely determined by the relative proportion of \ch{Fe^{2+}}/\ch{Fe^{3+}}, which is modulated by multiple competing physical, dynamical, and chemical processes \citep{lichtenberg_redox_2021, gaillard_redox_2022, schaefer_ferric_2024, frost_oxygen_1991}. We represent the mantle redox state with an oxygen fugacity $f\ch{O2}$, which is analogous to the partial pressure of \ch{O2} under ideal conditions \citep{bower_atmodeller_2025}. Since $f\ch{O2}$ primarily serves as a proxy for the oxidation state of Fe, we calculate it relative to the equivalent $f\mathrm{O^{IW}_2}$ of the iron-w\"ustite reaction \citep{fischer_equation_2011}, which depends on the surface temperature $T_\mathrm{surf}$ throughout the simulated evolution. The relative offset between the calculated mantle redox state and the iron-w\"ustite buffer parametrisation is an input parameter to PROTEUS simulations: $\Delta\IW = \log_{10}[f\ch{O2}/f\mathrm{O^{IW}_2}]$ \citep{sastre_geophysical_2026}. 

The total mass inventories of volatile elements (C-H-O-N-S) in the planet's atmosphere-mantle system decrease during the simulated evolution due to parametrised atmospheric escape processes \citep{postolec_atmospheric_2026}. Hydrodynamic energy-limited escape \citep{hunten_escape_1987, owen_review_2019} is included as non-fractionating atmospheric mass loss process \citep{nicholls_escape_2025, postolec_atmospheric_2026}, subject to an evolving stellar X-ray and ultraviolet spectrum through the MORS module \footnote{\url{https://proteus-framework.org/MORS/}} \citep{johnstone_active_2021, behr_muscles_2023, spada_radius_2013}. Orbital parameters are held constant during each simulation.

The vertical atmosphere climate calculations are performed with AGNI\footnote{\url{https://www.h-nicholls.space/AGNI/}}: a radiative-convective atmosphere model  applicable to diverse planetary regimes \citep{nicholls_agni_2025, nicholls_convective_2025}. The plane-parallel atmosphere is compositionally homogeneous with gas-phase mixing ratios set by degassing at the magma ocean surface \citep{lichtenberg_vertically_2021, elkins_linked_2008}. At each PROTEUS time-step, AGNI uses convective and radiative energy fluxes to calculate an energy-conserving solution for the 1D climate state \citep{nicholls_beyond_2026}. Atmospheric dry convection is modelled using a mixing-length formalism \citep{joyce_mlt_2023, prandtl_mlt_1925, marley_review_2015}. Radiative fluxes are calculated using the SOCRATES radiative transfer suite, which implements a two-stream method with Rayleigh scattering \citep{zdunkowski_twostream_1980, edwards_studies_1996, manners_fast_2024}. Gas opacities are represented and combined using correlated-$k$ equivalent extinction with 16 longwave and shortwave spectral bands \citep{lacis_description_1991, amundsen_treatment_2017}. Assuming that all modelled planet scenarios are tidally locked, we adopt a stellar zenith angle of $54.74^\circ$ and geometric scale factor of \sfrac{1}{4} \citep{hamano_emergence_2013, lebrun_thermal_2013, cronin_choice_2014}. The total upward-directed energy flux $F_\mathrm{atm}$ calculated by AGNI determines the simulated planets' rate of cooling, and thus, whether they solidify or maintain permanent magma oceans \citep{nicholls_redox_2024, lichtenberg_vertically_2021}. 

In previous studies, PROTEUS adopted the SPIDER\footnote{\url{https://proteus-framework.org/SPIDER/}} mantle dynamics model to simulate the combined physics of magma ocean energy transport, cooling, and mantle crystallisation \citep{bower_numerical_2018}. SPIDER uses 1D mixing-length theory to estimate mantle convective heat fluxes, enabling a 1D representation of the mantle thermo-compositional structure \citep{abe_thermal_1993, abe_early_1986, monteux_cooling_2016}. However, SPIDER is numerically unstable in the sub-Neptune regime because their internal conditions exceed the calibration domain of our empirical thermodynamic tables \citep{wolf_eos_2018, boley_fizzy_2023}. We appeal to boundary-layer theory to parametrise mantle dynamics and solidification, which remains a common approach in the literature for its simplicity and numerical stability \citep{schaefer_magma_2016, nicholls_chili_2026, krissansen-totton_predictions_2022, lebrun_thermal_2013, hamano_emergence_2013}. This choice enables testing of evolutionary exoplanet-lifetime retrievals for prototypical sub-Neptune exoplanets. In brief, we assume solidification of the magma ocean from the core-mantle boundary upwards \citep{maurice_onset_2017, solomatov_nonfractional_1993}, in which the volumetric mantle melt fraction evolves subject to its potential temperature $T_\mathrm{pot}$, initialised at 3200\,K to ensure a fully-molten initial state. The mantle is homogeneously heated by radioactive decay and latent heat released by crystallisation of its \ch{MgSiO3} material \citep{elkins_linked_2008}. Interior structure includes a metallic core, whose radius fraction $r_c=R_\mathrm{core}/R_\mathrm{int}$ is a simulation input parameter. Core density and mantle radius $R_\mathrm{int}$ are calculated at $t=0$ from the total planet mass and its material properties \citep{noack_parameteris_2020, nicholls_beyond_2026}. Appendix~\ref{app:boundary} details our interior structure calculation and mantle dynamics model \citep{calder_k218b_2026}, based upon \citet{schaefer_magma_2016}.

\subsection{Evolutionary retrievals within the PROTEUS framework}
\label{sec:methods_bo}

\subsubsection{Asynchronous Bayesian optimisation}

We estimate planets' interior and past properties, based on some observables, as a parameter retrieval problem. Retrieval methods aim to minimise the difference between  the simulated values and `true' values of some set of `observable' quantities. These methods estimate the best-fitting values on the non-observable input parameters which correspond to the best-fitting combination of simulated observables. The set of parameters which generate the best-fitting simulated observables provide insight into planets' internal structures, bulk and atmospheric compositions, and the multiple physical processes which we cannot directly observe \citep{madhusudhan_retrieval_2009, barstow_outstanding_2020}.  

The PROTEUS framework includes a modular forward model to simulate planetary evolution, and modes for dispatching the model. Several components of the modelled system cannot be expressed analytically. For example, gas opacities are derived from complex quantum mechanical calculations that generate pre-tabulated line-lists. Our radiative transfer calculations therefore rely on pre-computed tables of opacity coefficients, accessed via the FORTRAN-based SOCRATES code \citep{edwards_studies_1996}, which are non-linear functions of temperature, pressure, and composition \citep{grimm_database_2021, tennyson_exomol_2018}. The gradient of the objective function $f(x)$ is not accessible because the simulated physical system is not modelled through a wholly analytical formulation. Auto-differentiation approaches can be considered for obtaining gradients of $f(x)$ in other applications, but this requires special consideration from the initial stages of model construction and coding \citep{press_numerical_2007, julialang}. Numerical methods to approximate function gradients are computationally expensive, so a gradient-free optimisation approach to parameter retrieval is necessary. We strive to perform only point-wise evaluations to obtain a best-fitting set of some parameters. 

Batched or parallel Bayesian optimisation (BO) offers a principled framework to efficiently explore high-dimensional parameter spaces without requiring function gradients \citep{ginsbourger2010kriging,mockus_on_2005, madhusudhan_retrieval_2009, foreman_emcee_2013}. Bayesian optimisation provides an iterative gradient-free framework designed for expensive `black-box' objective functions \citep{jones_efficient_1998,garnett2023bayesian, savel_peering_2025}. In order to optimise the difference between simulated values and measurements made by observations with only point-wise evaluations, BO tracks the uncertainty on an objective function via a probabilistic surrogate model; in our case, a Gaussian process (GP) \citep{rasmussen_gaussian_2005}. This surrogate informs an acquisition function, proposing the next query location in the parameter space \citep{wilson2018maximizing,aigrain_gp_2023}.

Here, we incorporate asynchronous Bayesian optimisation (ABO) into the PROTEUS framework to efficiently dispatch the simulations required for fitting some observables. The input parameters to PROTEUS are categorised as fixed $\theta \in \Theta$ and inferred $x \in \mathcal{X} \subset \mathbb{R}^d$. It is $x$ which we will optimise for (retrieve) while $\theta$ are kept fixed throughout each retrieval. The forward model PROTEUS outputs of interest are collected in $\Gamma \subset \mathbb{R}^m$, allowing us to construct the deterministic mapping $\mathcal{M}_\theta: \mathcal{X} \mapsto \Gamma$. The function $\mathcal{M}_\theta$ represents a single forward evaluation of PROTEUS, to yield $\Gamma$. Given observables $\gamma_o \in \Gamma$, the goal of the optimisation is then to infer the values of parameters $x^* \in \mathcal{X}$ which satisfy the condition $\gamma_o = \mathcal{M}_\theta(x^*)$.

The outputs from PROTEUS simulations are used to evaluate a scalar objective function which quantifies deviation between simulated and `true' values of some observable quantities. The objective function encoding this optimality condition is a dimension-wise scaled Euclidean distance between the true values of the observables, $\gamma_o$, and the values calculated by PROTEUS,
\begin{equation}
    f(x) = -\log_{10}\Bigg[ \Big\|\mathbf{1}_m- \frac{\mathcal{M}_\theta(x)}{\gamma_o+\varepsilon}  \Big\|^2 + \varepsilon\Bigg],
    \label{eq:objective}
\end{equation}
where the fraction denominator is an element-wise normalisation factor, which acts as a regulariser for observables that will vary across different scales. We define $\mathbf{1}_m = (1, 1, \ ..., 1)^T \in \mathbb{R}^m$, such that larger values of $f$ represent better fits to the true observables. A log-transform is applied to the objective function $f$ for Gaussianity, which provides numerical stability within the optimisation scheme \citep{rasmussen_gaussian_2005}. A small tolerance term $\varepsilon=10^{-10}$ ensures the log-transform domain is well-defined, so larger $f$ represent better fitting scenarios, and $f=+10$ is a perfect fit. We also apply a log-transform to strictly positive-valued quantities that naturally scale logarithmically, rather than linearly; e.g, atmospheric mixing ratios and surface pressures.

Some of our results are presented with objective function values normalised to some best-fitting case, $f'=100f/f_\mathrm{best}$, such that $f'=100$ represents the corresponding best-fitting solution. Note that $f'$ remains a logarithmically-scaled quantity, so $f'=100$ represents a more than twice-better fitting parameter combination than $f'=50$.

\begin{figure}
    \centering
        \begin{tikzpicture}[
            >={Stealth[length=3mm]},
            lane/.style={draw=none, fill=none},
            tracklabel/.style={anchor=west, align=left},
            segbusy/.style={fill=blue!90!black!60!white!90},
            segidle/.style={fill=red!90!black!50!white!90},
            query/.style={draw=black, line width=0.4pt, dash pattern=on 2pt off 1.5pt},
            axis/.style={-Stealth, thick}
        ]
        \def\L{6}       
        \def\laneH{1.9}   
        \def\rowH{0.45}   
        \def\barH{0.10}   
        \def\gX{1.5}      
        \def\gY{0.6}      
        \def\barHbusy{0.10}   
        \def\barHidle{0.03}
        \def\labelX{\gX-1.0}   
        \def\headsep{0.5}     
        \def\sepH{0.30}   
        \def\busyLineW{0.8pt}                 
        \def\busyDash{on 3pt off 2pt}         
        \tikzset{
          busyline/.style={
            draw=blue!60!white,
            line width=0.8pt,
            dash pattern=on 3pt off 2pt,
            line cap=round
          }
        }
        \newcommand{\busyline}[3]{%
          \draw[busyline] (#2, #1) -- ++(#3, 0);
        }
        \node[lane, minimum width=\L cm, minimum height=\laneH cm, anchor=west] (sync)  at (\gX, \gY+\laneH+0.4) {};
        \node[lane, minimum width=\L cm, minimum height=\laneH cm, anchor=west] (async) at (\gX, \gY) {};
        \newcommand{\barLLH}[5]{%
          \path[#5, draw=none, rounded corners=0.3pt]
            (#2, {#1-0.5*#4}) rectangle ++(#3, #4);
        }
        \newcommand{\barBusy}[3]{\barLLH{#1}{#2}{#3}{\barHbusy}{segbusy}}
        \newcommand{\barIdle}[3]{\barLLH{#1}{#2}{#3}{\barHidle}{segidle}}
        \newcommand{\barLL}[4]{%
          \path[#4, draw=none, rounded corners=0.3pt]
            (#2, {#1-0.5*\barH}) rectangle ++(#3, \barH);
        }
        \newcommand{\barLR}[4]{%
          \path[#4, draw=none, rounded corners=0.3pt]
            (#2, {#1-0.5*\barH}) rectangle (#3, {#1+0.5*\barH});
        }
        \newcommand{\sepmark}[3][\sepH]{%
          \draw[query] (#2, {#3-0.5*#1}) -- (#2, {#3+0.5*#1});
        }
        \newcommand{\seprows}[3]{%
          \draw[query] (#1, {#3-0.5*\barH}) -- (#1, {#2+0.5*\barH});
        }
        \def\sa{ \gY+\laneH+0.4 + \laneH - 0.75*\rowH}
        \def\sb{ \gY+\laneH+0.4 + \laneH - 1.75*\rowH}
        \def\sc{ \gY+\laneH+0.4 + \laneH - 2.75*\rowH}
        \def\ya{ \gY + \laneH - 0.75*\rowH}
        \def\yb{ \gY + \laneH - 1.75*\rowH}
        \def\yc{ \gY + \laneH - 2.75*\rowH}
        \node[tracklabel] at (\labelX, \sa+\headsep) {Synchronous};
        \node[tracklabel] at (\labelX, \ya+\headsep) {Asynchronous};
        \node[tracklabel] at (\labelX, \sa) {$w_1$};
        \node[tracklabel] at (\labelX, \sb) {$w_2$};
        \node[tracklabel] at (\labelX, \sc) {$w_3$};
        \node[tracklabel] at (\labelX, \ya) {$w_1$};
        \node[tracklabel] at (\labelX, \yb) {$w_2$};
        \node[tracklabel] at (\labelX, \yc) {$w_3$};
        \barLL{\sa}{\gX+0.1}{1.9}{segbusy}
        \barIdle{\sa}{\gX+1.6+0.4}{0.6}
        \barLL{\sa}{\gX+2.6}{3.5}{segbusy}
        \busyline{\sa}{\gX+5.9}{0.6}
        \barLL{\sb}{\gX+0.1}{1.3}{segbusy}
        \barIdle{\sb}{\gX+1.4}{1.2}
        \barLL{\sb}{\gX+2.6}{1.6}{segbusy}
        \barIdle{\sb}{\gX+4.2}{1.4}
        \barLL{\sb}{\gX+5.6}{0.5}{segbusy}
        \busyline{\sb}{\gX+5.9}{0.6}
        \barLL{\sc}{\gX+0.1}{2.5}{segbusy}
        \barLL{\sc}{\gX+2.6}{2.2}{segbusy}
        \barIdle{\sc}{\gX+4.8}{0.8}
        \barLL{\sc}{\gX+5.6}{0.5}{segbusy}
        \busyline{\sc}{\gX+5.9}{0.6}
        \seprows{\gX+2.6}{\sa+0.2}{\sc-0.2}
        \seprows{\gX+5.6}{\sa+0.2}{\sc-0.2}
        \barLL{\ya}{\gX+0.1}{6}{segbusy}
        \sepmark{\gX+2.0}{\ya}
        \sepmark{\gX+5.0}{\ya}
        \busyline{\ya}{\gX+5.9}{0.6}
        \barLL{\yb}{\gX+0.1}{6}{segbusy}  
        \sepmark{\gX+1.4}{\yb}
        \sepmark{\gX+4}{\yb}
        \sepmark{\gX+5.3}{\yb}
        \busyline{\yb}{\gX+5.9}{0.6}
        \barLL{\yc}{\gX+0.1}{6}{segbusy}  
        \sepmark{\gX+2.6}{\yc}
        \sepmark{\gX+3.8}{\yc}
        \busyline{\yc}{\gX+5.9}{0.6}
        \draw[axis] (\gX+0.1, 0.6) -- ++(\L+0.2,0) node[below, yshift=-3pt] {Time};
        \coordinate (legendCenter) at ({\gX + 0.5*(\L+0.2)}, -0.45);
        \begin{scope}[shift={(legendCenter)}]
          \node[segbusy, draw=none, rounded corners=0.3pt,
                minimum width=0.7cm, minimum height=0.18cm] at (-3.6, 0) {};
          \node[anchor=west] at (-3.0, 0) {busy};
          \node[segidle, draw=none, rounded corners=0.3pt,
                minimum width=0.7cm, minimum height=0.18cm] at (-0.5, 0) {};
          \node[anchor=west] at (0.1, 0) {idle};
          \draw[query] (2.1, -0.15) -- (2.1, 0.15); 
          \node[anchor=west] at (2.4, 0) {query};
        \end{scope}
        \end{tikzpicture}   
    \caption{ Illustration of algorithmic dispatch of a forward-model under synchronous (\textbf{top}) and asynchronous (\textbf{bottom}) Bayesian optimisation approaches, for three worker CPUs. The asynchronous framework allows for more queries across the same wall-clock duration, by avoiding idle time. Illustration adapted from \citet{riegler2026standard}.}
    \label{fig:algorithm}
\end{figure}
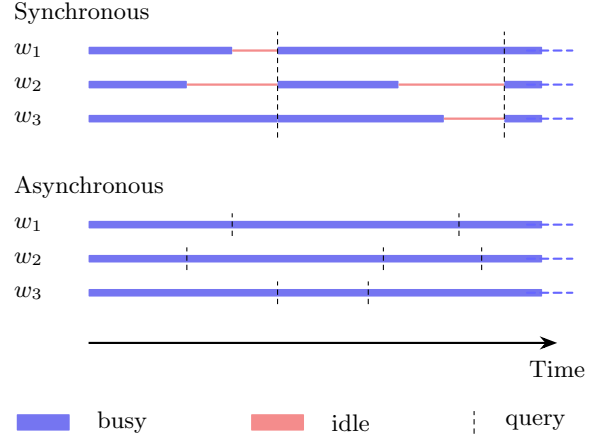

Each evaluation of our objective requires a PROTEUS forward simulation, which is computationally expensive (Appendix~\ref{app:evaltime}). The wall-clock duration of each ABO step is the sum of: (a) the initialisation time for the PROTEUS simulation, which involves loading Python and thermodynamic data; (b) the simulation's integration time, which scales with the numerical expense of the modelled physics; and (c) evaluation of $f$ in Equation~\ref{eq:objective} and determination of the next best $x$.  Importantly, since PROTEUS simulation wall-clock times are variable, dispatching batches of simulations synchronously would result in idle compute resources (Figure~\ref{fig:algorithm}, \citet{riegler2026standard}). Instead, our asynchronous batch BO routine allows us to maximally leverage high performance computing resources and permits PROTEUS simulations to run fully in parallel, while informing the same retrieval problem. This BO approach reduces the wall-clock time required for a complete parameter retrieval by minimising computer idle-time \citep{de2021asynchronous}. 

Pointwise evaluations of $f$ give a finite data set $D_n = (X,y)$, of inputs $X = (x_i)_{i=1}^n \in \mathbb{R}^{n \times d}$ and outputs $y = (y_i)_{i=1}^n \in \mathbb{R}^n$. To reason about the uncertainty in $f$ stemming from the finiteness of our data set, we place a zero-mean Gaussian process prior on the objective function. That is, we assume
\begin{align}
    f \sim \mathcal{GP}(0, k_\ell(\cdot, \cdot)),
\end{align}
under a positive-definite covariance function $k_\ell:\mathcal{X}^2 \mapsto \mathbb{R}_{>0}$, with hyperparameters $\ell$. Coupled with a Gaussian likelihood for the outputs, 
\begin{align}
    p(y \mid f, X) =  \mathcal{N}(y; f(X), \sigma_y^2 I_n),
\end{align}
with observation noise $\sigma_y$, this gives the function space Gaussian process posterior
\begin{align}
    \label{eq:gp_post}
    f\mid D_n \sim \mathcal{GP}(\mu_n(\cdot), \sigma^2_n(\cdot)).
\end{align}
For some input, $x \in \mathcal{X}$, the posterior mean and variance functions are given by
\begin{align}
    \mu_n(x) &=  k_\ell(x, X)(K_n + \sigma_y^2I_n)^{-1}y \\
    \sigma_n^2(x) &= k_\ell(x, x) - k_\ell(x, X)(K_n + \sigma_y^2I_n)^{-1}k_\ell(X, x),  
\end{align}
with $K_n \in \mathbb{R}^{n \times n}$ \citep{rasmussen_gaussian_2005}, the kernel matrix given by
\begin{equation}
    (K_n)_{ij} = k_\ell(x_i,x_j) \ \forall \ i, j \in [n].
\end{equation}
The model in \cref{eq:gp_post} and the query decision (the `acquisition function') give an iterative procedure for exploring the search space, $\mathcal{X}$, one point-wise evaluation of $f$ at a time. The ABO algorithm is formally described in Appendix~\ref{app:algo}.

Iterative Bayesian optimisation requires an initial guess to bootstrap the algorithm. Each of our asynchronous BO retrievals are initialised with three PROTEUS simulations (i.e., three evaluations of $f$) with parameters $x$ selected by random uniform sampling within specified parameter bounds. Each ABO retrieval's random number generator is seeded identically, so repeated retrievals use identical initial guesses, given the same set of parameters $x$, their priors, and choice of acquisition function and kernel.

Our baseline retrieval configuration uses a LogEI acquisition function (Section~\ref{sec:methods_acqf}), a Mat\'ern Gaussian process kernel with $\nu=3/2$ smoothing (Section~\ref{sec:methods_kernel}), five simultaneous worker processes (each allocated one CPU core), and a maximum of 100 PROTEUS evaluations after three randomly initialised bootstrapping samples. These choices represent `median' configuration scenarios, from the options currently implemented, and a modest resource allocation appropriate for this proof-of-concept study. We explore the scaling behaviour of our ABO algorithm's configuration and numerical parameters in Section~\ref{sec:results_schemes}. 

\subsubsection{Acquisition functions}
\label{sec:methods_acqf}

Using the surrogate model of the objective function in \cref{eq:gp_post}, we can make an uncertainty informed trade-off between exploration and exploitation on the search space $\mathcal{X}$. In particular, given data, $D_n$, the next query point (refer to Appendix~\ref{app:algo}), $x' \in \mathcal{X}$, is selected by
\begin{equation}
    \label{eq:acquire}
    x' = \arg \max_{x \in \mathcal{X}} \alpha(x \mid D_n),
\end{equation}
where $\alpha : \mathcal{X} \mapsto \mathbb{R}$ is a heuristic for the utility of querying a new input location in the parameter space, termed the `acquisition function'. 

Following \citet{riegler2026standard}, we consider only single-point standard acquisition functions. Three such `classical' standard acquisition functions, which are analytical and do not depend on Monte-Carlo sampling of the Gaussian process surrogate model constructed from PROTEUS simulations, are described below. We explore the relative performance of these different acquisition functions in Section~\ref{sec:results_schemes}.

Firstly, the Upper Confidence Bound (UCB),
\begin{equation}
    \alpha_{UCB}(x \mid D_n) = \mu_n(X) + \sqrt{\beta} \sigma_n(x),
\end{equation}
with exploration parameter $\beta=2$. UCB favours a suggested $x$ if either the GP predicts a good fit there (exploitation) or the uncertainty is high (exploration). More precisely: the UCB function is a simple heuristic based on optimising a quantile of the credibility interval, for example the 95\% outcome, representing the weighted sum of the posterior mean and standard deviation of the joint posterior function \citep{srinivas_information_2012}. The exploration parameter $\beta$ governs the compromise between exploration of the wider parameter space and exploitation of previously charted regions, informed by the current surrogate model.

Secondly, the logarithmic Probability of Improvement (LogPI) acquisition function,
\begin{equation}
     \alpha_{PI}(x \mid D_n) = \mathbb{P}_{f\mid D_n}[f(x) \ge \tau],
\end{equation}
with threshold $\tau$. LogPI selects the point $x$ most likely to improve upon $\tau$, without weighting by the magnitude of that improvement \citep{jones_efficient_1998}. Similarly to $\beta$ in the UCB, the choice of threshold governs the criterion's exploration-exploitation trade-off. Here, the threshold for selecting a new $x$ is chosen as $\tau =y_{i^*}$, with $i^* = \arg \max_{i}(y_i)$, representing the best-fitting scenario previously seen during the retrieval. We refer to \citet{jones_efficient_1998} for an in-depth discussion on LogPI threshold selection.

Thirdly, we consider the logarithmic Expected Improvement (LogEI) acquisition function,
\begin{equation}
    \alpha_{EI}(x \mid D_n) = \mathbb{E}_{f\mid D_n}[\max(f(x)- y_{i^*}, 0)],
\end{equation}
\citep{jones_efficient_1998, mockus_on_2005}. This function develops upon LogPI by additionally considering how much $f$ is improved under a candidate $x$. Adopting the log-transformed form offers improved numerical stability \citep{ament_unexpected_2023}.

\subsubsection{Gaussian process kernels}
\label{sec:methods_kernel}

Similarly to the choice of acquisition function, the most appropriate choice of Gaussian process kernel for this application is a non-trivial design decision. The Gaussian process is constructed and refined by running PROTEUS simulations during an ABO retrieval, so the kernel must analytically and functionally represent the behaviour of PROTEUS' modelled physics. We describe four different kernel parameter choices below, and assess their relative scaling performance in Section~\ref{sec:results_schemes}.

The radial basis function (RBF) is our simplest end-member option for the covariance function adopted as the Gaussian process kernel:
\begin{equation}
    \label{eq:rbf}
    k_\ell(x_i,x_j)  = \exp \Big( \frac{-d^2(x_i,x_j)}{2l^2} \Big)
\end{equation}
where $l$ is the non-dimensional length scale of the system (Section~\ref{sec:methods_acqf}). The RBF takes the form of a decaying squared exponential function, which is infinitely differentiable and therefore presumes that the objective function $f$ is a smooth function of the parameters $x$. The RBF generates Gaussian processes which may struggle to represent sensitive underlying physics \citep{stein_interpolati_1999}.

The Mat\'ern kernel generalises the RBF with additional terms,
\begin{equation}
    \label{eq:matern}
    k_\ell(x_i, x_j) = \frac{1}{\Gamma(\nu) 2^{\nu-1}} \Big[ \frac{\sqrt{2\nu}}{l} d(x_i, x_j) \Big]^\nu K_{\nu}\Big( \frac{\sqrt{2\nu}}{l} d(x_i, x_j) \Big),
\end{equation}
where the smoothness parameter $\nu$ permits more complex behaviour of the objective function (and thus, in our PROTEUS forward model). $\Gamma$ is the gamma function and $K_\nu$ is the modified Bessel function of the second kind \citep{rasmussen_gaussian_2005}. The choice of $\nu$ is debated in the literature, but it should be informed by the behaviour of the forward model \citep{stein_interpolati_1999}. The Mat\'ern kernel resolves to the RBF kernel (Equation~\ref{eq:rbf}) in the limit of $\nu\rightarrow\infty$, and to a decaying exponential in the case of $\nu=1/2$. We explore $\nu=[1/2\mathrm{,\,\,}3/2\mathrm{,\,\,}5/2]$ in Section~\ref{sec:results_schemes}, denoted as `$\mathrm{Mat}\nu$'.

For both kernels, the length-scales are chosen by optimising the density $p(l| {y})$, as
\begin{equation}
    l = \arg \max\limits_{l'} \ \log p({y}\mid l') + \log p(l'),
\end{equation}
with $p(y \mid l)$ the marginal likelihood and $p(l)$ a prior on length-scales. 

Following \citet{hvarfner_vanilla_2024} on Bayesian optimisation in high dimensions, we use 
\begin{equation}
    p ( l ) = \mathcal{LN} ( \sqrt{2} + \ln \sqrt{d} , \sqrt{3} ),
\end{equation}
a Log-normal distribution with dimension-scaled mean and variance.

\subsection{Input parameters and output variables}
\label{sec:methods_params}

PROTEUS has input and output variables. We divide these variables into four conceptual classes for the purposes of demonstrating the efficacy of evolutionary retrievals. 

There are two classes of input \textit{parameter} variables, whose values are not necessarily accessible from telescope observations of exoplanets. There remain strong scientific motivations for quantifying these parameters because they trace important physical processes and exoplanets' deep interior conditions \citep{lichtenberg_constraining_2025}. 

\subsubsection{Class\,P1 parameters}

Class\,P1 variables are input parameters that are held fixed for each of the exoplanet scenarios considered here: current planet age, orbital semi-major axis, and planet mass. Planet ages can be associated with the estimated age of their host-stars \citep{kippenhahn_stellar_2012}, which are observationally accessible through elemental compositions via nucleocosmochronometry \citep{soderblom_starage_2010, fowler_cosmochronology_1960}, spin-luminosity evolution \citep{baraffe_new_2015, jorgensen_Determina_2005}, and asteroseismology \citep{jendreieck_asteroseismology_2010, aerts_probing_2021, rauer_plato_2025}. Masses and orbital solutions are commonly estimated from radial velocity and transit-timing measurements \citep{seager_textbook_2011}. In the notation of Section \ref{sec:methods_bo}, these are fixed parameters $\theta$.

\subsubsection{Class\,P2 parameters}

Class\,P2 variables are input parameters that are to be estimated by our retrieval scheme, given the constraining observables. These quantities are summarised in Table~\ref{tab:param_range} alongside the adopted prior boundaries.

These quantities are inaccessible to remote sensing and must be estimated using models \citep{lichtenberg_constraining_2025, bloot_exoplanet_2023}. The metallic core radius fraction parameter, $r_c$, defines the interior structure. Core-building materials are denser than mantle-building materials, so varying $r_c$ from 0 to 1 leads to increased planet densities and decreased $R_\mathrm{obs}$, for a given total mass $M_p$ \citep{lodders_planetary_1998, noack_parameteris_2020, bloot_exoplanet_2023}. PROTEUS will retrieve $r_c$ as a Class\,P2 parameter. In the notation of Section \ref{sec:methods_bo}, these are input parameters $x$.

Multiple physical processes determine the redox conditions within planetary interiors, which regulate the partitioning and speciation of secondary atmospheres \citep{lichtenberg_redox_2021, frost_redox_2008, cottrell_earths_2025}. The Solar System planets exhibit diverse redox conditions \citep{sossi_physicochem_2025}, which are also expected to vary spatially and temporally within the interiors of specific bodies \citep{maurice_redox_2023, hirschmann_iw_2021, kress_redox_1991}. Metallic core segregation is intimately linked to mantle redox conditions, since Fe becomes increasingly incompatible in silicate melts under reducing conditions \citep{wade_core_2005, tronnes_core_2019}. Following convention, we adopt the surface fugacity of oxygen $f\ch{O2}$ as a proxy for mantle redox, quantifying it as the logarithmic oxygen fugacity offset relative to the iron-w\"ustite buffer reaction: $\Delta\IW$ \citep{frost_oxygen_1991, oneill_the_2002}. The oxygen fugacity $\Delta \IW$ is adopted as a Class\,P2 parameter to be estimated.

Three additional Class\,P2 parameters define a planet's initial inventory of volatile elements, endowed following formation and early boil off \citep{raymond_planet_2022, nicholls_redox_2024, tang_cpml_2024, lichtenberg_water_2019, krijt_chemical_2023}. We vary the total initial hydrogen content, quantified as parts-per-million relative to the mantle mass $\mathrm{H_{ppmw}}$. Carbon and sulfur budgets are defined relative to hydrogen: the bulk C/H mass ratio, and the bulk S/H mass ratio. These three quantities scale with total planet mass $M_p$.  \citet{wang_elements_2018} present $\mathrm{H_{ppmw}}=109$, $\mathrm{C/H}=1.0$, and $\mathrm{S/H}=2.2$ as concordant estimates on Earth's early volatile inventory, which each carry substantial uncertainties \citep{peslier_water_2017, dreibus_core_1996, sossi_physicochem_2025}.

\subsubsection{Class\,O1 observables}

There are also two classes of \textit{output} variables calculated by PROTEUS simulations. Class\,O1 variables are output quantities used to evaluate the objective function (Equation~\ref{eq:objective}) as the difference between these observed quantities' ground-truth values and the outputs of each PROTEUS simulation. Careful selection of these constraining observables is important; they must be accessible to remote sensing methods aimed at real exoplanets, while remaining physically correlated with the estimated Class\,P2 parameters in order for them to act as effective constraints. We choose seven observable variables to constrain the demonstration retrievals presented in this work. 

Photospheric radii $R_\mathrm{obs}$ can be measured using transmission spectroscopy and photometry \citep{seager_textbook_2011, carter_jwst_2022}. We derive $R_\mathrm{obs}$ from the hydrostatic atmosphere solution's 20\,mbar pressure level, appropriate for mid-infrared exoplanet limb measurements \citep{lopez_understanding_2014, fortney_transit_2005, nicholls_beyond_2026}. The structure of molecular absorption and emission features probe atmospheric scale heights, related to the photospheric composition and structure: $H_p=RT/(\mu g)$, where $R$ is the ideal gas constant \citep{ito_theoretical_2015, Wakeford_Transmissio_2015, barstow_transit_2015}. We adopt upper-atmosphere temperature, gravitational acceleration, and molecular weight as Class\,O1 observables \citep{molliere_interp_2022, birkby_spectro_2018, nicholls_beyond_2026}. Furthermore, spectroscopic features attributed to specific molecular species enable estimates on atmospheric metallicity ratios, which we also adopt: C/O, S/O, and O/H \citep{childs_composition_2023, madhusudhan_co_2012, welbanks_massmetall_2019, savel_precise_2026}. 

These seven observables represent a compromise between instrumental limitations and necessary retrieval constraints. In reality, these would be derived jointly from spectroscopy or photometry and have individual uncertainties. The observables would be derived from substantial processing and reduction of raw measurement data: steps which would occur before the application of PROTEUS retrievals to a real exoplanet \citep{gordon_jwst_2022}. Other quantities could also be considered as constraints; e.g., day-night brightness temperature contrasts \citep{kreidberg_absence_2019, hammond_linking_2017}. Our ABO retrievals do not consider uncertainties on these observables, since they are derived from synthetic prototype exoplanets, but comparison against real observations would necessitate systematic incorporation of observables' uncertainties in selecting $x$ and evaluating $f$ \citep{davey_investigati_2025, hayes_optimizing_2020}. 

\subsubsection{Class\,O2 quantities}

Lastly, we record a large number of simulation output variables that are not accessible by remote sensing methods (Class\,O2). These outputs are computed as a result of the fully-coupled nature of our multi-physics simulator. Examples of these include: mantle melt fraction $\Phi_m$, surface temperature and pressure conditions, vertical atmospheric structure, planet bulk density, tidal heat flux, and the Bond albedo. Estimating these quantities is scientifically motivated as we ultimately strive to interpret exoplanets' deep interior conditions, habitability, and provenance.

\subsection{Prototypical exoplanets as case-studies}
\label{sec:methods_cases}

This study presents a proof-of-concept for efficiently retrieving the lifetime histories of exoplanets using Bayesian statistics. To remain agnostic of observational uncertainties associated with specific real exoplanets, we generate three ground-truth baseline scenarios to represent some hypothetical observations. These are informed by populations of exoplanets which are emerging from ongoing surveys: 
\begin{itemize}
    \item (SN) A young, temperate `Sub-Neptune' exoplanet.
    \item (SE) An older, warm `Super-Earth' exoplanet.
    \item (TR) A hot, young `Terrestrial' Earth-sized exoplanet.
\end{itemize}
These are three representative prototypes of the wider exoplanet population, which enable us to study the situations under which evolutionary retrievals are feasible. The three prototypical scenarios are differentiated by their Class\,P1 and P2 input parameters. We consider differences in three fixed parameters (Class\,P1): orbital semi-major axis, current planet age, and initial planet mass.  These prototypes could readily be replaced with real planets that have had their properties constrained by observations.

The semi-major axis is held fixed and an eccentricity $e=0$ is adopted for each scenario \citep{nicholls_tidal_2025}. We consider a host star of mass $0.273 M_\odot$, modelled after L\,98-59 \citep{demangeon_warm_2021, cadieux_l9859_2025}. This M3-type star is a quiet M-dwarf that has been well-studied and for which a semi-synthetic emission spectrum is available from the MegaMUSCLES database \citep{behr_muscles_2023}. The stellar spectrum evolves bolometrically and spectroscopically during the simulations, self-consistently with the planet \citep{johnstone_active_2021}. In application to real exoplanets, PROTEUS' evolutionary retrievals could adopt stellar parameters appropriate to each specific scenario \citep{johnstone_active_2021, baraffe_new_2015, luger_extreme_2015}. 

We adopt an initial planet age of 50\,Myr, relative to the host star age, which represents an early initial state after the stellar nebular and protoplanetary disk have dissipated \citep{kippenhahn_stellar_2012, deeg_formation_2018, valencia_diversity_2025}. Thus, simulation integration times $t_\mathrm{evo}$ can be interpreted as present-day planet ages with a 50\,Myr offset. These models neglect early accretion processes, atmospheric boil off after nebula dissipation, potential giant impacts, and orbital migration \citep{tang_cpml_2024, rogers_unveiling_2021, schlichting_impact_2018, kimura_formation_2020, chyba_delivery_1990}.

These three prototypes also differ by the Class\,P2 parameters in Table~\ref{tab:var_defs}~and~\ref{tab:param_range}, which are to be retrieved by our ABO retrievals (Section~\ref{sec:methods_bo}).

\subsection{Static snapshot retrievals with InferAGNI}
\label{sec:methods_static}

Exoplanets' current interior conditions can be inferentially estimated from observations using static non-evolving models \citep{valencia_diversity_2025, Zeng2019, dorn_can_2015}. InferAGNI represents a leading tool for inferring planets' present-day internal and atmospheric structures, and already incorporates a physically comprehensive radiative-convective-chemical atmosphere model into its calculations \citep{nicholls_agni_2025, nicholls_beyond_2026}.  We directly compare our ABO evolutionary retrievals against InferAGNI static retrievals, using the same three prototypical exoplanet cases. 

InferAGNI performs parameter retrieval using an affine-invariant Markov-chain Monte Carlo (MCMC) method through the `emcee' Python package \citep{foreman_emcee_2013, gilks_markov_1996, goodman_ensemble_2010}. The MCMC log-likelihood function is evaluated using a regular grid linear interpolator forward model, constructed upon a pre-computed library of 504000 diverse exoplanet scenarios \citep{nicholls_beyond_2026}. These scenarios agnostically span a range of planet masses, envelope- and core-fractions, metallicities, instellation fluxes, and host stars. Radial static structures are solved in 1D with a binary interior (metallic core and mantle) and overlying atmosphere. The interior equation of state assumes a fully-differentiated iron core and silicate \ch{MgSiO3} silicate mantle, and is applicable for planet masses up to $50M_\oplus$ \citep{seager_mass_2007}. The atmosphere is solved using AGNI --- the same atmosphere model applied in PROTEUS simulations \citep{nicholls_agni_2025, nicholls_convective_2025}. The temperature structure is obtained self-consistently with the vapour mixing ratios by applying FastChem at  thermochemical equilibrium \citep{kitzmann_fastchem_2024, stock_fastchem_2018}. The MCMC algorithm performs 8000 steps across 14 simultaneous workers, before a thinning factor of 4 is applied to the flattened MCMC chains, and the final 5\% of samples are selected for comparison against PROTEUS' ABO. 

Controlled comparison between ABO and InferAGNI is enabled by their adoption of the same atmosphere model (AGNI) and conceptually equivalent constraints (Section~\ref{sec:results_static}). All of the Class\,O1 observables used to constrain ABO (Section~\ref{sec:methods_params}) have analogous or identical quantities within the InferAGNI framework. Similarly, all of the estimated Class\,P2 parameters can be mapped from InferAGNI's MCMC posteriors. Given their common ground-truth scenarios (Section~\ref{sec:methods_cases}), differences in the two retrieval frameworks' inferred planet structures and compositions reflect the efficacy an evolutionary retrieval.

\section{Results} 
\label{sec:results}

\subsection{Establishing a ground-truth}
\label{sec:results_truth}

\begin{figure}
    \centering
    \includegraphics[width=0.94\linewidth]{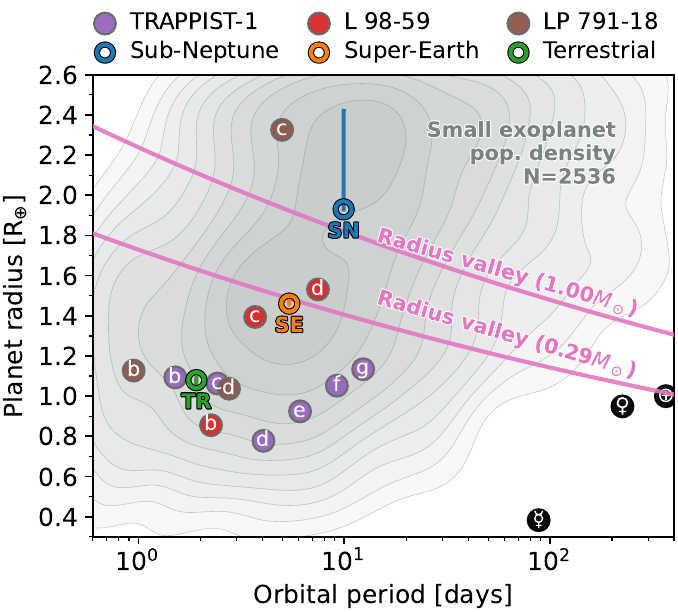}
    \caption{Simulated radius-period evolution of our three prototypical planets (blue, green, orange lines), adopted as `ground-truth' scenarios. Grey contours visualise the Gaussian-kernel population density of the surveyed exoplanet population, drawn from \url{exoplanet.eu}. The pink lines show empirical fits to the small-exoplanet radius valley, for a Solar-mass star and an L\,98-59-mass star \citep{ho_deep_2023}.}
    \label{fig:exoplanets}
\end{figure}

Firstly, we run three standalone PROTEUS simulations to establish ground-truth scenarios for our prototypical exoplanet cases. The blue, green, and orange points in Figure \ref{fig:exoplanets} show the evolving radii $R_\mathrm{obs}$ calculated from these simulations, versus the planets' fixed orbital periods. These simulations are contextualised by the radius-period distribution of the surveyed exoplanet population, drawn from \url{exoplanet.eu}, which is visualised by kernel-density estimate contours. 

The simulated photospheric radii of the ground-truth planets (blue, orange, green lines) are initially inflated due to their hot interior conditions, far from radiative equilibrium. This exposes them to large escape rates with large outgoing energy fluxes \citep{lopez_born_2017, rogers_most_2015, owen_evap_2017}. Thermal radiation emission to space, balanced against stellar irradiation, causes their interiors to cool and partially crystallise. Their atmospheres contract according to the modelled atmospheric thermodynamics. Their calculated atmospheric compositions evolve self-consistently with their thermal evolution and mass loss (Section~\ref{sec:methods_proteus}) --- Figure~\ref{fig:truthevo} plots the thermal and compositional evolution of these three simulations. None of the cases completely solidify; they terminate with $\Phi_m^\mathrm{SN}=63\%$, $\Phi_m^\mathrm{SE}=82\%$, and $\Phi_m^\mathrm{TR}=30\%$. Permanent magma oceans are capable of efficiently sequestering their volatiles against atmospheric escape processes  \citep{DornLichtenberg2021, hirschmann_magma_2012} and driving changes in atmospheric metallicity ratios \citep{nicholls_escape_2025, cesario_reflation_2026}. Overall, three prototypical cases act as representative members of their corresponding exoplanet populations. We attempt to retrieve their parameters using the PROTEUS-derived observables in this proof-of-concept study. The SN and SE scenarios simultaneously probe the physics of hydrodynamic mass-loss and interior-atmosphere partitioning, in the sensitive regime around the radius valley \citep{heng_gradient_2025,ito_hydrodynami_2021, rogers_road_2025}. The TR case exhibits stark compositional evolution, testing how atmospheric constraints may alleviate mass-radius degeneracies via evolutionary modelling.

The SN case (blue line in Figure~\ref{fig:exoplanets}) moves through a densely populated region of the phase space corresponding to the broader sub-Neptune population \citep{rogers_road_2025, valencia_diversity_2025}. We adopt a geochemically reduced interior for this planet (Table~\ref{tab:var_defs}) with a large initial hydrogen inventory, so this prototype could be categorised as a `gas dwarf' type sub-Neptune with a \ch{H2}-dominated atmosphere (Figure~\ref{fig:truthevo}; \citet{calder_most_2026}). In reality, the sub-Neptune population may include a diversity of planetary structures and compositions \citep{lichtenberg_constraining_2025}. Figure~\ref{fig:exoplanets} demonstrates that the parameters chosen to define our SN prototype are appropriate to represent the wider sub-Neptune population. Importantly, while its radius evolves from $\sim2.4$~to~$\sim1.9R_\oplus$, the planet remains above the radius valley after 1\,Gyr of integration time (pink lines in Figure~\ref{fig:exoplanets}; \cite{ho_deep_2023}).

Leading theories suggest that super-Earths arise from sub-Neptune progenitors through multiple processes which conspire to remove their atmospheres and decrease their radii \citep{Fulton_RadVal_2017, david_cks_2021, rogers_most_2015}. These long-term processes overprint a range of potential post-formation scenarios; e.g., formation interior or exterior to the ice and soot lines of their protoplanetary disks \citep{li_soot_2025, bergin_exoplanet_2023, krijt_chemical_2023, lichtenberg_water_2019, kimura_formation_2020, Venturini2020, Burn2024}. The super-Earths population may comprise multiple sub-families of planet types \citep{cherubim_oxidation_2025, heng_gradient_2025}. Our SE case is initialised with an intermediately oxidised and metal-rich composition, and simulated over an extended integration time of 2\,Gyr (Table~\ref{tab:param_range}). Its atmosphere transitions from \ch{CO}-dominated to \ch{CH4}-dominated, while \ch{H2O} remains the subdominant  component (Figure~\ref{fig:truthevo}), driven by thermochemistry as the surface cools and the preferential loss of carbon relative to hydrogen from the atmosphere \citep{nicholls_escape_2025, cesario_reflation_2026, dixon_co2_1997, sossi_solubility_2023}. Our SE case experiences a smaller radius change than the SN case because of the abundance of heavier elements (C and O) which generate smaller atmospheric scale heights, although it remains susceptible to complete volatile loss. The SE evolves through a highly-populated region of radius-period space below the radius valley, so it can act as a representative of the super-Earth population \citep{rogers_most_2015, ginzburg_super-earth_2016, ho_shallower_2024}.  

The Terrestrial planet case (green) undergoes minimal absolute changes in its radius $R_\mathrm{obs}$, since its mantle is initialised with oxidised conditions --- comparable to Earth's \citep{frost_redox_2008, nicklas_redox_2018} --- that generate high molecular weight atmospheres with small atmospheric scale heights (Figure~\ref{fig:exoplanets}). The TR case's increased X-ray irradiation, compared to the SN and SE cases, cause  hydrodynamic removal of the majority of its initial carbon inventory within the short 100\,Myr integration time (Figure~\ref{fig:truthevo}). Its atmosphere transitions from \ch{CO2}- to \ch{H2O}-dominated through this interval, and there is a notable increase in \ch{H2S} abundance \citep{bower_retention_2022}. This third scenario would represent a challenging target for transmission spectroscopy with current observatories; its final photospheric radius $1.07R_\oplus$ is small and largely set by the planet's interior component $R_\mathrm{int}/R_\mathrm{obs}=92\%$ \citep{seager_textbook_2011, barstow_outstanding_2020}. The compositional observables will be key constraints on its parameters for PROTEUS' evolutionary ABO retrievals (Table~\ref{tab:var_defs}).

\begin{table*}
\centering
\begin{tabularx}{\linewidth}{lllll}
    \hline
    (\textbf{P1}) Fixed parameter     & Sub-Neptune   & Super-Earth  & Terrestrial  & Accessibility  \\ 
    \hline
    Current planet age & 1000     & 2000   & 100 Myr   & Stellar dating    \\
    Semimajor axis     & 0.06     & 0.04   & 0.02 AU   & Radial velocity, transit timing    \\
    Initial planet mass    & 3.0      & 1.9    & 1.0 $M_\oplus$ & Radial velocity, transit timing \\

    \hline 
    (\textbf{P2}) Estimated parameter  & Sub-Neptune   & Super-Earth  & Terrestrial  & Accessibility \\ 
    \hline 

    Metallic core fraction [\% radius] & 35       & 40     & 55\% radius    & \textit{Models and inference only}   \\
    Mantle oxygen fugacity (i.e., redox state) & $\IW-3.0$  & $\IW+0.5$& $\IW+3.0$  & \textit{Models and inference only}  \\
    Initial bulk H inventory ($\Hppmw$)   & 10000      & 10000    & 2000 ppmw & \textit{Models and inference only} \\
    Initial bulk C inventory (C/H) & 0.5      & 1.5    & 2.5   & \textit{Models and inference only}     \\
    Initial bulk S inventory (S/H)  & 0.8      & 2.0    & 0.8   & \textit{Models and inference only}   \\
    
    \hline
    (\textbf{O1}) Observable variable's truth value  & Sub-Neptune   & Super-Earth  & Terrestrial  & Accessibility \\ 
    \hline
    Final photosphere radius        & 1.92  & 1.45   & 1.07 $R_\oplus$ & Trans. spectroscopy \\
    Final photosphere temperature   & 313   & 409    & 582 K & Trans. / Emit. spectroscopy \\
    Final photosphere gravity       & 8.09  & 9.11   & 8.58 m/s & Trans.  spectroscopy \\
    Final atmosphere molec. weight  & 4.1   & 18.9   & 25.6 g/mol & Trans. / Emit. spectroscopy \\
    Final atmosphere C/O  (wt.)      & 3.19  & 0.87   & 0.19 & Trans. / Emit. spectroscopy \\
    Final atmosphere S/O  (wt.)      & 1.07  & 0.57   & 0.12 & Trans. / Emit. spectroscopy \\
    Final atmosphere O/H  (wt.)      & 0.16  & 2.80   & 13.04 & Trans. / Emit. spectroscopy \\
\end{tabularx}
\caption{Three prototype exoplanet cases defining our ground-truths differ by their input parameters and synthetic observables. The fixed input parameters (Class\,P1) are held constant in each case, while the estimated parameters (Class\,P2) are retrieved by optimising calculated observables (Class\,O1) against ground-truth values. Fixed parameters and observables are observationally accessible; estimated parameters require modelling to constrain.}
\label{tab:var_defs}
\end{table*}

\begin{table}
    \centering
    \begin{tabularx}{\linewidth}{llll}
    \hline
    \textbf{(P2)} Parameter & Symbol         & Minimum   & Maximum   \\ 
    \hline 
    Metallic core frac. & $r_c$          & 30       & 70\%       \\
    Mantle redox state  & $\Delta\IW$          & -4.0        & +4.0        \\
    Initial bulk H inventory & $\mathrm{H_{ppmw}}$ & 1000       & 20\,000   \\
    Initial bulk C inventory & C/H               & 0.1       & 4.0      \\
    Initial bulk S inventory & S/H               & 0.1       & 4.0      \\
    \end{tabularx}
    \caption{Class\,P2 parameters' minimum and maximum prior bounds. These quantities are estimated by Bayesian optimisation against ground-truth values on the Class\,O1 synthetic observable variables in Table~\ref{tab:var_defs}.}
    \label{tab:param_range}
\end{table}

\subsection{Retrieval via Bayesian optimisation}
\label{sec:results_bo}

Evolutionary retrievals are demonstrably feasible with asynchronous Bayesian optimisation. Overall, we find that PROTEUS' ability to fit the observables with ABO --- maximising $f$ --- depends on the leading-order physics active within each planetary regime. Figures~\ref{fig:obs_converge} and \ref{fig:par_converge} summarise these retrievals. Our results show that PROTEUS' retrievals can converge upon the synthetic ground-truth observables despite an `expensive' planetary evolution forward model. 

Figure~\ref{fig:obs_converge} plots the evolutionary scenarios explored by PROTEUS during its ABO retrievals, for our three prototype exoplanets (columns). The first seven rows of coloured panels (rows a-g) show PROTEUS-simulated time evolution of the constraining Class\,O1 observables. Each line represents a single invocation of PROTEUS, with the best-fitting scenario that maximises the objective function $f$ plotted in black. The bottom three rows of Figure~\ref{fig:obs_converge} (rows h-j) are Class\,O2 output variables calculated by PROTEUS, which are not observationally accessible and do not factor into $f$. Ground-truth values are shown by unfilled circular markers. In parallel, Figure~\ref{fig:par_converge} presents the Class\,P2 parameter combinations where PROTEUS dispatches evolutionary simulations (forward models) during its ABO retrievals (y-axes), versus the corresponding normalised objective function value $f'$ (x-axis). Note that $f'$ is a logarithmically-scaled measure for the goodness of fit. Square markers in Figure~\ref{fig:par_converge} indicate ground-truth  Class\,P2 parameters, analogous to circular markers in Figure~\ref{fig:obs_converge}.  The well-performing cases with accurately retrieved parameters and observables are those where our best-fitting scenario (black) converges upon the ground-truth (unfilled markers). Table~\ref{tab:best_fit} quantifies best-fitting values and their relative linear errors.

In the favourable physical regime of our TR case, evolutionary retrievals enable inference of its complete histories and provide deep insight into its interior conditions (Figure~\ref{fig:obs_converge} column C). However, less favourable physical regimes (our SN and SE cases; Figure~\ref{fig:par_converge} columns A and B) can make optimisation of certain parameters difficult, while facing the same degeneracies as the existing static-structure retrieval paradigm \citep{dorn_can_2015, Zeng2019, seager_mass_2007, huang_magrathea_2022}. We present each exoplanet in subsections below.

\subsubsection{Sub-Neptune case}

PROTEUS' ABO retrieval is able to reproduce several of the SN case's ground-truth observable values with good accuracy. Under the best-fitting scenario,  $f=-0.35$. For example, a small error of 5.9\% is obtained on atmospheric S/O, which enables estimation of the planet's physically-correlated initial S/H budget (Figure~\ref{fig:par_converge}Ae). The evolutionary retrieval of this SN prototype is able to infer the planet's initial sulfur budget, by resolving the control S/H has over its present-day atmospheric S/O ratio. 

The SN's retrieved $R_\mathrm{obs}$ is 21.1\% below the ground-truth value (Figure~\ref{fig:obs_converge}Aa). The under-estimated radius value leads to an over-estimated core mass fraction and under-estimated initial volatile budget (Figure~\ref{fig:par_converge}Aa/Ac). The core-building material is denser than the mantle-building material, and smaller volatile budgets outgas thinner atmospheres, which yields smaller photospheric radii. The SN ground-truth is configured with a reducing mantle ($\Delta\IW=-3$) and large initial volatile inventory ($\mathrm{H_{ppmw}}=10^4$). This geochemical regime outgasses \ch{H2}-dominated atmospheres with small molecular weights and large scale heights, regardless of the surface and interior thermal conditions. Insensitivity between the compositionally-related Class\,O1 observables ($\mu_\mathrm{obs}$, C/O, O/H) and the esimtaed Class\,P2 parameters ($\Delta \IW$, $\Hppmw$, C/H) inhibits accurate parameter estimation.

Misfit on the SN's photospheric temperature, $T_\mathrm{obs}$, is small (0.2\%; Figure~\ref{fig:obs_converge}Ab). A good fit is obtained because upper-atmospheric temperature structure is primarily determined by the bolometric irradiation flux, which is decoupled from the Class\,P2 parameters varied here, except through their control over Rayleigh scattering. Future incorporation of orbital migration physics would lead to variations in $T_\mathrm{obs}$ as a function of inferred Class\,P2 parameters. 

\subsubsection{Super-Earth case}

The super-Earth retrieval terminates with a closer best-fit to the ground-truth ($f=-0.19$). This planet exists within a regime strongly shaped by atmospheric escape --- in the adopted energy-limited regime, atmospheric mass loss scales as $\propto R_\mathrm{XUV}^3 F_\mathrm{XUV}$. The planet's radius evolves throughout the simulated evolution (Figure~\ref{fig:exoplanets}) and creates a large cross-section for X-ray absorption to drive photoevaporation, so complete atmospheric erosion arises in some of the PROTEUS simulations probed during the ABO retrieval (Figure~\ref{fig:obs_converge}Bi). The outgassed atmosphere is dominated by carbon-bearing species (Figure~\ref{fig:truthevo}d) which are poorly soluble in the planet's magma ocean, so carbon atoms dominate the hydrodynamically escaping outflow \citep{bower_retention_2022, dixon_co2_1997}.  Further evolution of this planet would drive continued changes in its observables, so simulation \textit{timing} --- corresponding to planet ages --- directly shapes the synthetic observables compared against the ground-truth constraints when evaluating the objective function (Equation~\ref{eq:objective}).

Differential interior-atmosphere partitioning between the volatile elements, alongside escape, drives evolution in observable $\mu_\mathrm{obs}$, S/O, C/O, and O/H (Figure~\ref{fig:obs_converge}Bc,e,f,g). With its mildly reducing redox state $\Delta\IW=+0.5$ and sensitivity to photoevaporative carbon depletion, compared to sulfur retention, the SE case accesses a diversity of atmospheric compositions. The Class\,O1 observables $\mu_\mathrm{obs}$ and S/O (misfits of 4.8\% and 6.7\%) enable PROTEUS to accurately estimate the SE ground-truth mantle's oxidation state as being mildly reducing ($\Delta\IW=+1.2$; Figure~\ref{fig:par_converge}Bb). 

Although the PROTEUS' ABO retrieval converges upon SE's ground-truth photospheric radius (8.3\% misfit; Figure~\ref{fig:obs_converge}Ba), it struggles to accurately infer the SE core fraction and initial volatile budget (Figure~\ref{fig:par_converge}Ba,c). This planet's propensity for atmosphere erosion biases evolutionary pathways towards small surface pressures (Figure~\ref{fig:obs_converge}Bi), so the retrieval favours low-density interiors with small metallic core fractions --- rather than extended atmospheres --- to obtain a good fit on the radius (Figure~\ref{fig:par_converge}Ba,c). Similarly to the SN case, this behaviour is a known degeneracy between atmosphere- and core-mass fractions, established by current static-structure approaches \citep{unterborn_pressure_2019, dorn_can_2015, Zeng2019}. We note that PROTEUS explores a narrow range of $\Hppmw$ during the SE retrieval case, near the Bayesian prior's lower bound and far below the ground-truth's value (Figure~\ref{fig:par_converge}Bc).

\subsubsection{Terrestrial case}

\begin{figure*}
    \centering
    \includegraphics[width=0.95\linewidth]{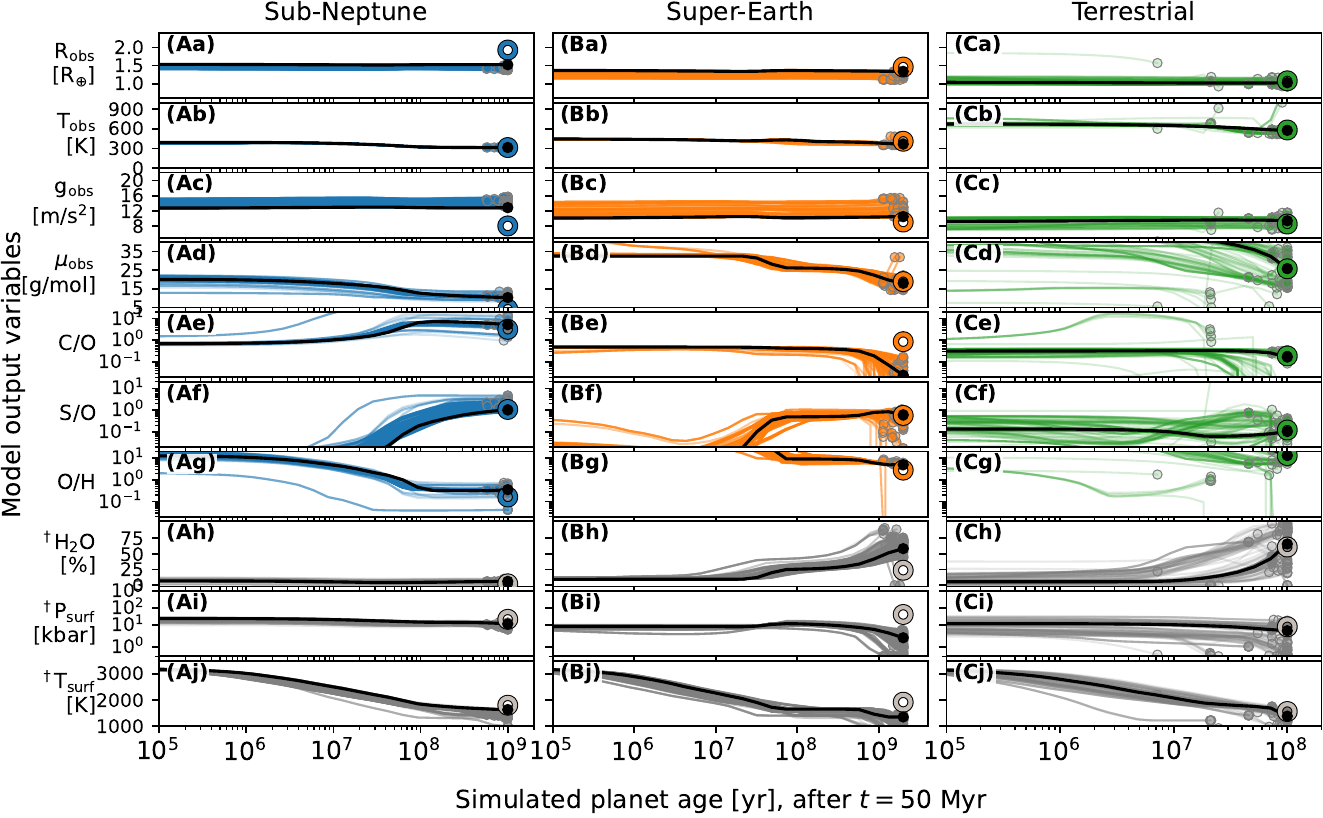}
    \caption{Simulated observable variables calculated by PROTEUS (rows) as a function of planet age (x-axes), for our three prototypical exoplanet scenarios (columns). Ground-truth observables are shown by unfilled circle markers in each panel. Class\,O1 observables (coloured rows) are used to constrain the retrieval, with ground-truth values shown by unfilled circles. Grey rows, labelled with $\dagger$, show Class\,O2 quantities not used in fitting. Black lines show best-fitting scenario. Values and errors are presented in Table~\ref{tab:best_fit}.}
    \label{fig:obs_converge}
    
    \includegraphics[width=0.95\linewidth]{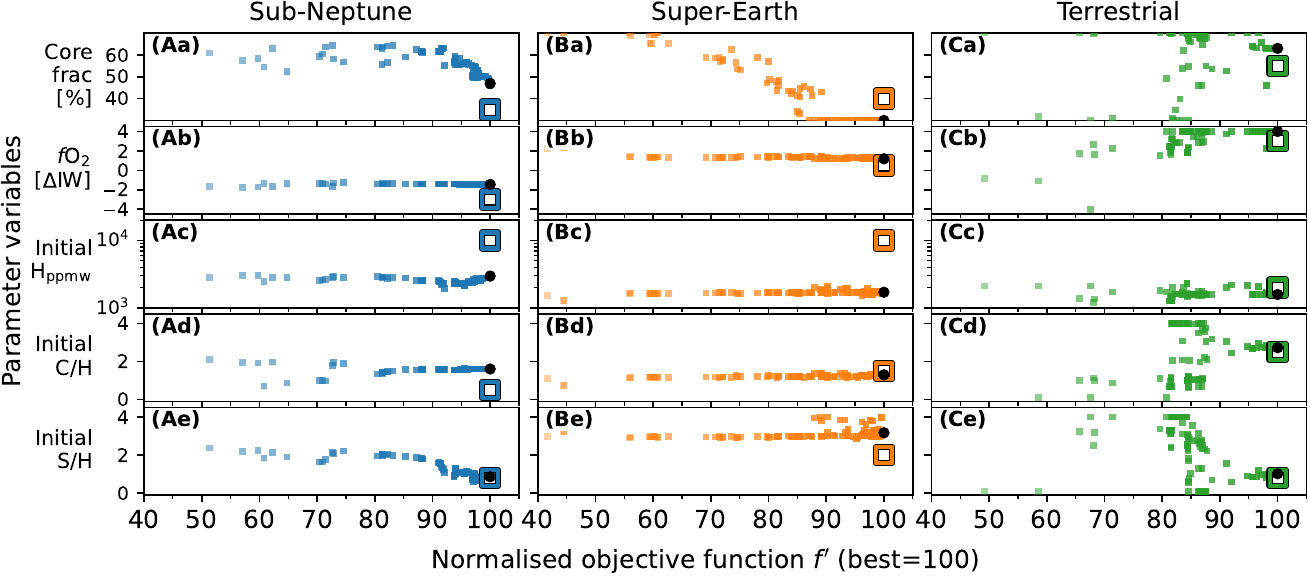}
    \caption{Parameter evaluation locations $x$ (y-axes) and their normalised objectives $f'$ value (x-axes) explored by PROTEUS' ABO retrievals, corresponding to the same simulations as Figure~\ref{fig:obs_converge}. Best-fitting solutions are indicated by the black points (Table~\ref{tab:best_fit}). Ground-truth values on the parameters (Table~\ref{tab:var_defs}) are shown by unfilled square markers at $f'=100$. Some parameters are better estimated than others, depending on the planet (columns and colours).}
    \label{fig:par_converge}
\end{figure*}

Our final scenario, a warm Terrestrial prototype, presents the best evolutionary retrieval performance ($f=+1.33$). Its adopted physical regime and properties enable favourable correlation between observables and parameters, so all the Class\,O1 observables are well fit: black lines converge upon the ground-truth values shown by unfilled green markers in Figure~\ref{fig:obs_converge} column C. Excellent correspondence between our best-fitting and ground-truth TR observables enables the accurate estimation of the TR prototype's hidden Class\,P2 parameters (black point in Figure~\ref{fig:par_converge} column C; Table~\ref{tab:best_fit}).

The TR case's surface conditions and atmospheric composition evolve together throughout the explored evolutionary pathways (Figure~\ref{fig:obs_converge}Ce-Cj). Elemental ratios evolve non-monotonically across multiple orders of magnitude: not resolvable by static models since the driving physical processes are time-dependent. Mantle solidification continually decreases the reservoir of molten material capable of efficiently storing \ch{H2O}, thereby degassing initial volatiles and exposing them to hydrodynamic escape \citep{nicholls_escape_2025, arora_thin_2026}. Atmospheric molecular weight decreases as its initially \ch{CO2}-dominated composition becomes diluted with degassed \ch{H2O} (Figure~\ref{fig:obs_converge}Cd,h; \citet{bower_retention_2022, cesario_reflation_2026}). The total surface pressure generally exhibits a net decrease over time, although many cases experience complete atmospheric erosion (Figure~\ref{fig:obs_converge}Ci). 

Of the observables probing atmospheric composition --- metallicity ratios and atmospheric molecular weight --- the largest misfit is in S/O (14.8\%) and smallest is in $\mu_\mathrm{obs}$ (0.6\%). Well-fit observables enable accurate estimates of the TR planet's initial volatile inventory; the best-fit scenario finds $\mathrm{C/H}=2.7$ and $\mathrm{H_{ppmw}}=1570$ compared to the ground-truth's 2.5 and 2000, respectively. Convergence on to 21\% misfit on $\Hppmw$ (1570 versus 2000\,ppmw initial bulk-planetary hydrogen content) is a small error; we anticipate wide diversity in the hydrogen budgets afforded between the exoplanets being observed --- depending on their provenance \citep{sossi_physicochem_2025, deeg_formation_2018}, early accretion \citep{kimura_water_2026, bitsch_water_2019}, and protracted exposure to atmospheric escape \citep{zahnle_evolution_1988, kite_atmosphere_2020} --- exemplified by the order-of-magnitude uncertainties in Earth's own present-day water budget \citep{peslier_water_2017}.

Accurate retrieval estimates of the TR planet's initial volatile budget strongly brackets the range of late-stage surface pressures compatible with observations, despite $P_\mathrm{surf}$ not being directly observable (Figure~\ref{fig:obs_converge}Ci). Alongside compositional measures, initial volatile content is jointly inferred from the radius constraint ($R_\mathrm{obs}$ misfit of 4.8\%); simultaneously modelling atmospheric physics and chemistry lifts the degeneracy on the bulk density faced by our SN and SE cases. The TR planet's best-fitting core fraction $r_c=63\%$ thereby obtains a small misfit 12.7\%, in contrast to $\sim25\%$ for the SN and SE scenarios (c.f., Table~\ref{tab:best_fit}). 

Spectroscopic observations estimate metallicity ratios and mean molecular weights more readily than the abundances of specific molecules \citep{madhusudhan_co_2012, parmentier_thermal_2018, soni_effect_2023, piaulet_evidence_2023}. Modelling connects these quantities. For example, a narrow range of \ch{H2O} mixing ratios are thermochemically compatible with a given O/H ratio \citep{kitzmann_fastchem_2024, taylor_a_2026}. Figure~\ref{fig:obs_converge}Ch shows that a range of \ch{H2O} volume mixing ratios can arise from the scenarios explored by PROTEUS' retrieval algorithm. Yet, the best-fitting case reproduces our ground-truth \ch{H2O} mixing ratio (61\% at the surface) with a misfit error of 7.6\%. The estimation of particular molecular abundances provides insight into exoplanet's climatic conditions and potential for habitability \citep{madhusudhan_chemical_2023, shorttle_k218b_2024, wogan_jwst_2024, benneke_toi270d_2024}.

\subsection{Evolutionary versus static retrieval approaches}
\label{sec:results_static}

We now directly compare our evolutionary-retrieval framework to the current static-retrieval paradigm, in their ability to infer exoplanets'  observationally-inaccessible interior conditions. The ground-truth scenarios have optically thick atmospheres which would shroud their surfaces from direct observation, so their interior phase states would only be inferentially characterisable. We use the planets' pressure-temperature ($P$-$T$) conditions as a phase space for comparing evolutionary (PROTEUS; Section~\ref{sec:methods_bo}) and static (InferAGNI; Section~\ref{sec:methods_static}) approaches. 

Figure~\ref{fig:static} plots the P-T conditions explored by both retrieval methods, when applied to each exoplanet prototype (panels). We reference these conditions against our ground-truth (unfilled markers) and contextualise them with mantle melting and \ch{H2}-\ch{H2O} miscibility boundaries (red-dashed and blue-dash-dotted lines, respectively). We flatten the MCMC sampler's walkers and derive posteriors from the final 5\% of MCMC samples (pink contours in Figure~\ref{fig:static}), which can be interpreted as being best-fitting static retrieval estimates of the three exoplanets' surface conditions. In comparison, greyscale lines show all 100 pathways of evolving P-T conditions simulated by PROTEUS during each ABO retrieval --- including poorly-fitting initial guesses (thinner lines), better fits (thicker lines), and the best-fitting scenarios (black lines). Note the inverted y-axis.

Overall, we find that evolutionary retrievals can provide greater physical insight into the surface conditions of sub-Neptune, super-Earth, and terrestrial-mass exoplanets. Modelled surface P-T conditions evolve past the ground-truth, while the static approach has low precision and can be biased towards cooler surfaces. We discuss each exoplanet case in the subsections below.

\subsubsection{Sub-Neptune case}

In the sub-Neptune case, InferAGNI retrieval posteriors and PROTEUS' simulated evolution pathways converge upon similar surface P-T conditions (Figure~\ref{fig:static}a). The static-retrieval's posterior on surface conditions ($T_\mathrm{surf}=1404^{+185}_{-150}$\,K; pink contours) probes P-T pairings comparable to the end states of PROTEUS simulations (grey points), for which  the best-fitting evolutionary simulation terminates with $T_\mathrm{surf}=1626$\,K, at 1\,Gyr (black point). Correspondence between the two approaches can be understood by considering the SN's parameter regime and our model choices. Firstly, following from Section~\ref{sec:results_bo}, the reducing conditions defining our SN prototype (Table~\ref{tab:var_defs}) generate atmospheres strongly dominated by \ch{H2} \citep{katyal_effect_2020, bower_atmodeller_2025}. Changes to surface temperature and pressure have little impact on the observable photospheric $R_\mathrm{obs}$ and $\mu_\mathrm{obs}$, allowing degenerate parameter combinations to produce observables consistent with the SN ground-truth. This is the same ongoing challenge ambiguating real sub-Neptunes' interior conditions from telescope observations \citep{madhusudhan_exoplanetary_2019, lichtenberg_constraining_2025}. Secondly, both retrieval  approaches incorporate the same radiative-convective-chemical code (AGNI) to simulate the planet's atmospheric structure, while being constrained by the same ground-truth observations, so similar surface conditions will necessarily generate the same Class\,O1 observables. 

The best-fitting evolutionary retrieval approximates the SN ground-truth surface conditions ($T_\mathrm{surf}=1797$\,K; blue circle) with better accuracy than the static-retrieval's posterior median: 9.5\% versus 21.8\% relative error on $T_\mathrm{surf}$, respectively. The InferAGNI algorithm explores cooler surface conditions and higher surface pressures  (pink contours in Figure~\ref{fig:static}a) than the PROTEUS-ABO approach, because PROTEUS models evolve over a range of large internal luminosities that arise from initially molten interiors --- equivalent to $T_\mathrm{int}>0$\,K \citep{noti_effects_2024, nicholls_convective_2025, fortney_unified_2008}. 

PROTEUS simulations model a suite of time-evolving processes which conspire to limit the set of physically permissible surface conditions within the wider P-T phase space (grey lines in Figure~\ref{fig:static}a and Figure~\ref{fig:obs_converge}Bj). The range of conditions spanned by these pathways is determined by variations in our retrieved Class\,P2 parameters (Table~\ref{tab:var_defs}). Importantly, these physical constraints mean that PROTEUS' ABO retrieval algorithm does not probe the lower $P_\mathrm{surf}<6000\mathrm{\,bar}$ conditions suggested by the static-structure retrieval posterior (pink contour), so the best-fitting evolutionary case sits within the same region of P-T space as the ground-truth (blue marker).  All SN cases remain immiscible above the \ch{H2}-\ch{H2O} demixing binodal \citep{howard_demix_2025}. 

The best-fitting SN scenario (black line; Figure~\ref{fig:static}a) does not terminate at the same P-T conditions as the ground truth. Instead, it evolves closely past the blue marker and overshoots towards cooler conditions with lower surface pressures. Atmospheric escape only marginally decreases the surface pressure during its 1\,Gyr evolution; changes in the SN's radius are driven by cooling and thermal contraction \citep{nicholls_escape_2025, tang_reassessing_2025, lopez_how_2012}. Had the best-fitting SN simulation terminated earlier --- at a point between the 115~and~220\,Myr PROTEUS time-steps --- it would have obtained strong correspondence to the ground-truth's P-T conditions (blue marker). 

\subsubsection{Super-Earth case}

Compared to the sub-Neptune case, both retrieval approaches provide relatively poor estimates for our super-Earth's surface conditions (Figure~\ref{fig:static}b). The SE prototype is intentionally constructed in a parameter regime where sensitive physical feedbacks make a diverse range of atmospheric compositions and structures possible (Figure~\ref{fig:truthevo}d; \citet{cherubim_oxidation_2025, gaillard_diverse_2021, bower_atmodeller_2025}). This planet is defined with the oldest age (longest integration time) and its final surface conditions sit on the silicate solidus, so physics provides only a narrow window for retaining thick atmospheres with specific compositions 2\,Gyr after its formation (Table~\ref{tab:var_defs}). 

The SE planet's exposure to competing physical processes does introduce parameter-observable relationships that can enable precise estimation on the inferred Class\,P2 parameters, such as its C/H ratio with a small 13.7\% misfit (Section~\ref{sec:discuss_bo}). However, Figure~\ref{fig:static}b demonstrates that both retrieval approaches are unable to converge upon the SE ground-truth's P-T conditions before exhausting their allocated computer resources. The best-fitting PROTEUS simulation overshoots towards the thinner atmospheres and cooler surfaces comparable to the static-structure median ($T_\mathrm{surf}=1346$\,K versus $1403^{+27}_{-16}$\,K, and $P_\mathrm{surf}=1946$\,bar versus $1267^{392}_{-155}$\,bar; Table~\ref{tab:best_fit}).  The ABO algorithm preferentially explores the Bayesian prior's lower-limit on initial volatile content ($\mathrm{H_{ppmw}}$ in Table~\ref{tab:param_range}), which is where the retrieval's bootstrapping samples were randomly initialised. The relative inaccessibility of the SE ground-truth's conditions to retrieval makes PROTEUS' ABO retrievals susceptible to the same mass-radius degeneracy inhibiting the static approach. 

\subsubsection{Terrestrial case}

Section~\ref{sec:results_bo} presented PROTEUS retrievals of the Terrestrial prototype --- representing our best performing case study. Its modelled observables evolve non-monotonically and are strongly correlated to the inferred Class\,P2 parameters, enabling insight into the physics actively shaping its final state and hidden interior conditions (Figure~\ref{fig:obs_converge} column C). 

\begin{figure}
    \centering
    \includegraphics[width=1\linewidth]{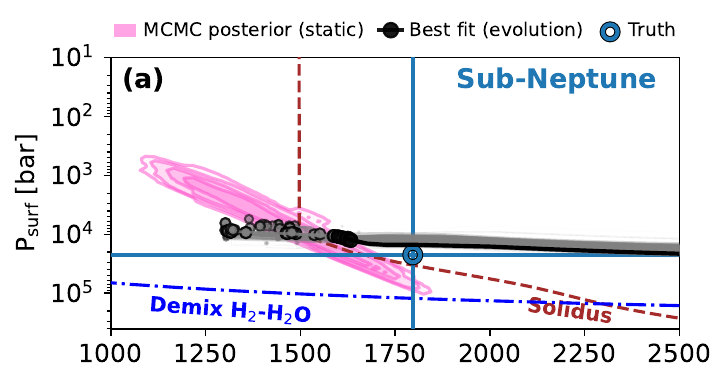}%
    \vspace*{-1mm}
    \includegraphics[width=1\linewidth]{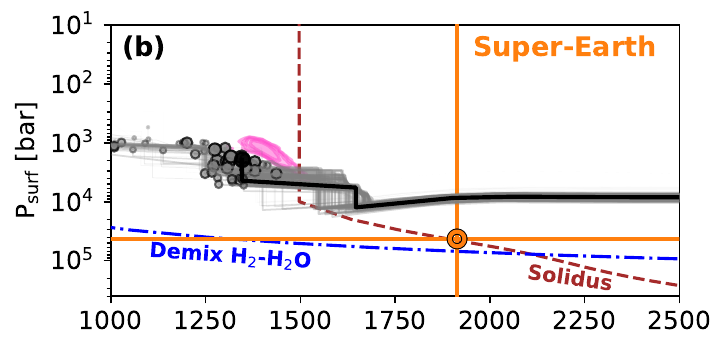}%
    \vspace*{-1mm}
    \includegraphics[width=1\linewidth]{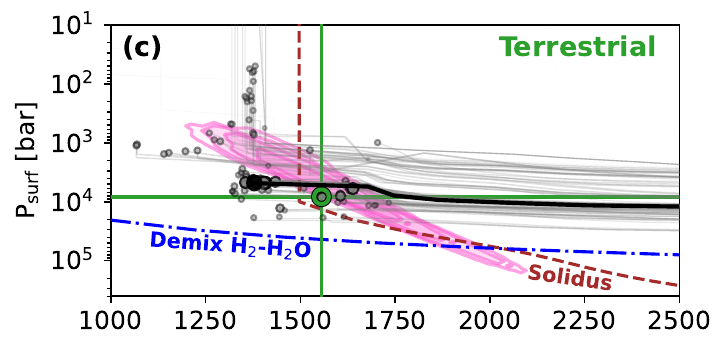}%
    \vspace*{-1mm}
    \caption{Surface pressure-temperature conditions explored through both static and evolutionary retrieval approaches, for three exoplanet cases (panels). Note the inverted y-axis. Pink contours show the posterior density of best-fitting sample points, from the final 5\% of the flattened MCMC chain of static-structure InferAGNI retrievals. Greyscale lines show individual PROTEUS simulations dispatched during evolutionary retrievals; endpoints are indicated with dots and best-fitting simulations are highlighted in black (Table~\ref{tab:best_fit}). The ground-truth $P_\mathrm{surf}$-$T_\mathrm{surf}$ are shown by blue, orange, and green vertical/horizontal lines and the unfilled markers. Dashed red line plots a silicate solidus, demarcating molten and solidified surfaces \citep{wolf_eos_2018}. Dash-dotted blue line shows an analytical fit for the \ch{H2}-\ch{H2O} de-mixing binodal, calculated using the \ch{H2O} mixing ratios of the ground-truth scenarios \citep{howard_demix_2025, Bergermann_demix_2024, Vlasov_demix_2023}.}
    \label{fig:static}
\end{figure}

Physics permits a wide range of surface pressures and temperatures for this planet, since it may readily lose its atmospheric blanketing effect through hydrodynamic escape \citep{hamano_emergence_2013, nicholls_chili_2026, ito_hydrodynami_2021}, whilst remaining strongly heated by stellar radiation on a short orbital period. Several exploratory PROTEUS simulations terminate with thinner atmospheres $P_\mathrm{surf}<90\mathrm{\,bar}$ and cooler surfaces $T_\mathrm{surf}<1250\mathrm{\,K}$, having evolved through a wide range of earlier states (grey lines in Figure~\ref{fig:static}c); PROTEUS appropriately disfavours these poorly-fitting scenarios. In comparison, InferAGNI's TR posterior spans multiple surface regimes, including those with permanent magma oceans (right of dashed red line) and immiscible \ch{H2}-\ch{H2O} conditions (below dash-dotted blue line).

However, the ground-truth's final surface temperature is 1558\,K, so a direct comparison of the best-fitting PROTEUS scenario's final surface temperature (1379\,K) would suggest it has worse performance in estimating $T_\mathrm{surf}$ than the static-retrieval approach (1560\,K posterior median). Note that the MCMC posterior on surface pressure has a large relative uncertainty that exceeds unity ($P_\mathrm{surf}=4792^{+9261}_{-3001}$\,bar; pink contour). A wide range of surface pressures are available to the InferAGNI retrieval as it jointly varies its atmosphere mass-fraction alongside the core mass-fraction parameters, to make up the same total radius $R_\mathrm{obs}$ and mass $M_p$. Earlier termination of the best-fitting evolutionary scenario (black line) --- at integration times between 73~and~99\,Myr --- would yield a better 1529\,K estimate for the ground-truth's surface temperature, so timing remains a key determiner of accurate P-T inference.

\subsection{Comparison of optimisation schemes}
\label{sec:results_schemes}

Tools used to interpret exoplanet observations strive for computational performance; enabling broader parameter exploration, comprehensive modelling of the relevant physics, and insight into population-level trends \citep{barstow_outstanding_2020, rotman_enabling_2025, fisher_HowDoWe_2022}. Yet, computational resources are limited, and many codes run on single CPU cores. Efficient retrieval algorithms mitigate this tension \citep{garvin_machine_2024, madhusudhan_retrieval_2009, foreman_emcee_2013}. It is unclear, \textit{a priori}, which algorithms and functions are best suited to this application of asynchronous Bayesian optimisation. Here, we consider three suites of acquisition functions, Gaussian process kernels, and distributions of CPU resources. We run a series of additional PROTEUS ABO retrievals where the baseline configuration is identical to the previous sections (Section~\ref{sec:methods_bo}). We adopt the same Terrestrial planet ground-truth Class\,O1 observables and inferential Class\,P2 parameters (Table~\ref{tab:var_defs}).

Figure~\ref{fig:perf} presents the scaling behaviours of PROTEUS' Bayesian retrievals, depending on the adopted retrieval configuration (line colours). Each panel plots the normalised objective function $f'$ (scatter points) calculated by each invocation of PROTEUS simulations, versus the index of that PROTEUS simulation during the particular retrieval instance. The log-transformed objective $f$ is normalised to the best-fitting case in each panel, $f'=100f/f_\mathrm{best}$, such that $f'=100$ represents the best-fitting solution and the values $f'<0$ are plotted at $f'=0$. Solid lines plot the best-seen $f'$ versus each evaluation number.

\subsubsection{Acquisition functions}

Figure~\ref{fig:perf}a presents retrieval scaling behaviour using three analytical acquisition functions (line and point colours). The retrievals with UCB and LogEI functions converge upon a similar best-fitting scenario. LogEI determines its best $f=+1.33$ after 68 PROTEUS evaluations, but then shows no further improvement before  the retrieval process terminates at 100 complete PROTEUS simulations (black line in Figure~\ref{fig:perf}). The UCB function marginally outperforms the LogEI baseline after 91 PROTEUS simulations (best $f=+1.38$, pink line). The LogPI configuration begins with a better initial guess from its bootstrapping, but it performs poorly during the retrieval and continuously explores badly fitting scenarios, even after $>90$ PROTEUS simulations (cyan line). 

Both LogEI and UCB require approximately 55 PROTEUS simulations for $f$ to be substantially increased beyond their initial guess (black and pink lines in Figure~\ref{fig:perf}a). Sensitivity tests which extended the LogEI case to 200~simulations showed no further improvement upon $f=+1.33$ (not shown). 

The LogEI acquisition function converges upon its best-fitting objective $f$ sooner than the UCB, ending with only a marginally worse $f'$; we conclude that LogEI is the preferred acquisition function for this task. We do not find it necessary to adopt a penalisation-based acquisition function since each acquisition is more informed than the previous one (Figure \ref{fig:algorithm}; \citet{riegler2026standard, rasmussen_gaussian_2005}).

\subsubsection{Gaussian process kernels}

Figure~\ref{fig:perf}b presents retrieval performing with the LogEI acquisition function, but considers four Gaussian process kernel configurations (Section~\ref{sec:methods_kernel}). Larger values of the Mat\'ern kernel smoothing parameter $\nu$ are able to resolve greater sensitives in the objective $f$ as a function of $x$. The Mat\'ern kernel is smoothest in the limit of $\nu\rightarrow\infty$, where is it equivalent to the RBF. 

We find that the RBF kernel has the worst retrieval performance (pink line in Figure~\ref{fig:perf}b); its best objective $f=-0.05$ is found after 13 iterations and does not improve before the 100 evaluations are completed. This result reflects Section~\ref{sec:results_bo}: the physics resolved by PROTEUS' planetary evolution simulations are particularly sensitive to the retrieved Class\,P2 parameters in the Terrestrial planet's regime, where a diverse range of oxidised atmospheres are simultaneously shaped by magma ocean outgassing and hydrodynamic escape. Previous studies have established that smooth Gaussian process kernels cannot capture sensitives inherent to complex physical processes \citep{stein_interpolati_1999}. 

Retrieval performance is not a monotonic function of the kernel smoothness: the RBF is equivalent to $\nu\rightarrow\infty$, yet the smoothest Mat\'ern kernel achieves the best objective $f=+1.42$ across all kernel configurations after 65 PROTEUS simulations ($\nu=5/2$, brown line). The roughest Mat\'ern configuration marginally improves upon our $\nu=3/2$ baseline, which has the best $f=+1.33$ after 68 simulations. Our intermediate-smoothness $\nu=3/2$ configuration presents performance intermediate to the other kernels, although it shows no improvement to $f$ after 43 simulations.

We conclude that a Mat\'ern covariance function with $\nu=5/2$ may be the most appropriate choice as a kernel for this task since it obtains the best objective $f'$ --- all else equal --- and presents reliable improvement as a function of simulation count (x-axis). However, modifications to the underlying PROTEUS code --- e.g., by incorporation of orbital dynamics --- may yield physical sensitivities that are better represented by a different functional form or smoothness parameter.

\subsubsection{Worker numbers}

\begin{figure}
    \centering
    \includegraphics[width=1\linewidth]{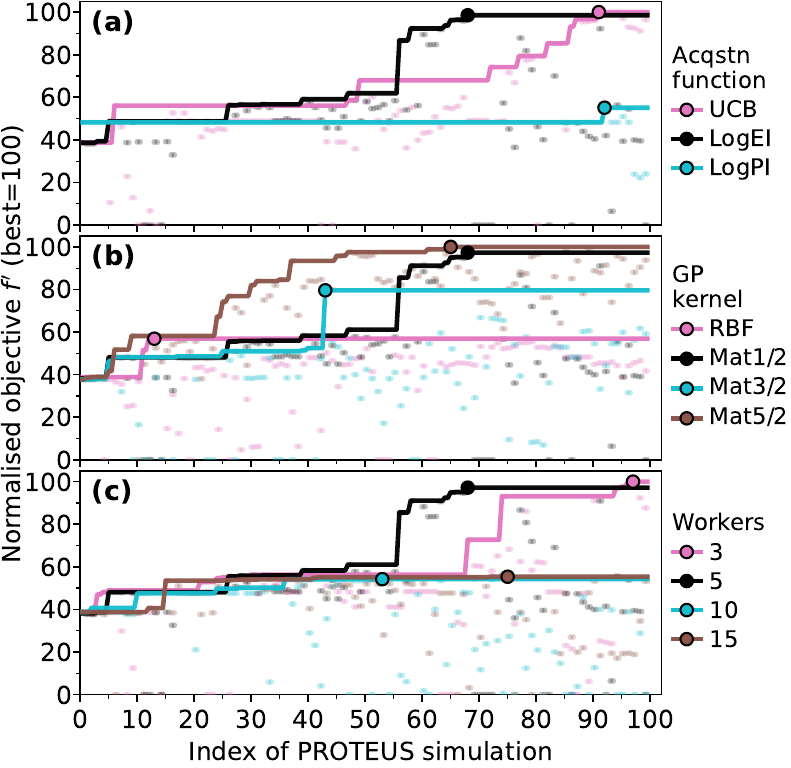}
    \caption{Scaling behaviours of PROTEUS' asynchronous Bayesian optimisation retrievals applied to our Terrestrial exoplanet prototype. Scatter points plot the normalised objective function $f'$ against the dispatched evaluation number (x-axis). Lines quantify the best-fitting $f'$ seen.  \textbf{Top}: three acquisition functions (Section~\ref{sec:methods_acqf}). \textbf{Middle}: four Gaussian process kernel functions (Section~\ref{sec:methods_kernel}). \textbf{Bottom}: four different numbers of parallel worker processes (3, 5, 10, and 15 CPU cores). The baseline retrieval configuration follows Section~\ref{sec:methods_bo}.}
    \label{fig:perf}
\end{figure}

Figure~\ref{fig:perf}c uses the LogEI acquisition function and Mat\'ern $\nu=3/2$ kernel from our baseline configuration, but considers different numbers of simultaneous workers. Our PROTEUS ABO retrievals are run with a pre-defined number of CPU cores, with one worker process per core. Each worker asynchronously dispatches a new PROTEUS simulation --- immediately after its previous simulation terminates --- informed by the acquisition function and the Gaussian process which has been constructed from previous PROTEUS simulations (Section~\ref{sec:methods_bo}). Regardless of worker allocations, each retrieval expires after 100 PROTEUS simulations have been performed. 

The choice of 3 workers marginally outperforms our baseline retrieval 5-worker configuration (best $f=+1.43$ versus $f=1.33$) after 97 PROTEUS simulations have run, although the 5-worker configuration finds its best $f$ sooner. In general, we find improved retrieval performance with fewer simultaneous workers; configurations with 10 or 15 workers show negligible improvement in $f$ after 53 PROTEUS simulations are performed and have poorer best-fitting objective values $f$ at 100 PROTEUS simulations. This behaviour arises from our BO dispatch scheme and fixed 100 simulation budget, since a smaller number of simultaneous workers allows each BO selection of parameters $x$ to follow from greater number of previous simulations, rather than being dispatched concurrently \citep{riegler2026standard, de2021asynchronous}. 

\section{Discussion} 
\label{sec:discuss}

\subsection{Retrievals across billions of years}
\label{sec:discuss_bo}

Exoplanets' initial and current states are accessible from present-day observations when parameter inference is approached as a retrieval problem that incorporates planetary evolution forward models. We have applied computationally expensive planetary evolution simulations to model the physics connecting planets' earliest conditions to their currently observed states. Our machine learning framework uses asynchronous Bayesian optimisation to dispatch these models, extending beyond the current static-snapshot paradigm. However, evolutionary retrieval performance and efficiency varies depending on the planetary and physical regimes considered.

Evolutionary retrieval of our SN prototype accurately constrains the planet's initial sulfur budget (our S/H parameter) by connection with an observed S/O ratio. Atmospheric S/O ratio is accessible through transmission spectroscopy targeting \ch{SO2}, \ch{CS2}, and \ch{H2S} absorption features \citep{panagiotou_sulfur_2026, dai_photochemic_2026, jordan_organosulfu_2026}. Atmospheric S/O is also physically correlated with mantle redox conditions, because the solubility of sulfide \ch{S^{2-}} ions in molten silicate materials is sensitive to mantle redox conditions (proxied by $\Delta\IW$; \citet{namur_mercury_2016}). Our SE retrieval demonstrates that mantle redox conditions can be inferred from these observables, by self-consistently incorporating thermochemistry and  interior-atmosphere volatile partitioning into forward models \citep{ito_coupling_2026, wogan_jwst_2024}. This finding stands in contrast to the common adoption of free-chemistry retrievals within the literature, which can preferentially infer physically infeasible combinations of chemical species \citep{madhusudhan_carbon-bearing_2023, al_a_2022, luu_volatile_2024, piaulet_evidence_2023}. Future applications of our evolutionary retrievals could enable PROTEUS' module for chemical kinetics (VULCAN; \cite{tsai_biogenic_2024, tsai_comparative_2021}), since disequilibrium photochemistry can imprint observable signatures \citep{panagiotou_sulfur_2026, dai_photochemic_2026, tsai_direct_2022, nicholls_temperaturechemistry_2023}. To enable direct comparisons against future JWST and ELT measurements, synthetic transmission spectra \citep{molliere_petitradtra_2019} could be derived from our radiative-convective atmosphere modelling --- accounting for aerosol complexities --- potentially with a $\chi^2$ formulation for $f$ \citep{Barstow_Unveilingc_2020, macdonald_trident_2022, fairman_the_2024, barstow_transit_2015}.

Our multi-physics PROTEUS forward model estimates values on planetary parameters which are not directly observable, including those inaccessible to current static-structure approaches. For example, on our $\Delta \IW$ Class\,P2 parameter, which represents the upper-mantle redox state of these simulated planets \citep{sastre_geophysical_2026}. Estimation of $\Delta \IW$ acts to probe the multiple simultaneously-acting processes which set the redox conditions of planetary interiors (Section~\ref{sec:methods_proteus}). Terrestrial-mass planets remain observationally difficult targets, in comparison to sub-Neptunes and super-Earths, because of their decreased planet-to-star size and flux ratios. Yet, good performance within the oxidised terrestrial planet regime may enable deep physical insight, when observables are accessible \citep{madhusudhan_exoplanetary_2016}. Estimates of Earth's present hydrogen and carbon budgets vary across two orders of magnitude \citep{peslier_water_2017, clark_carbon_1982}. So, placing accurate constraints on exoplanet's initial volatile inventories --- even to within an order of magnitude --- is sufficient for identifying specific formation scenarios; i.e., where this planet formed within the protoplanetary disk relative to the water ice-line and soot-line, and the efficiency of primordial envelope boil-off \citep{li_soot_2025, lichtenberg_water_2019, boitard_iceline_2025, krijt_chemical_2023, tang_cpml_2024, ginzburg_cmpl_2018}. 

We self-consistently model these planets' evolving atmospheric chemistry and vertical structures, so observational measures of composition jointly constrain the retrievals alongside the planets' observable bulk properties (e.g., their radii). This approach is demonstrated to lift degeneracies on exoplanets' interior structures and redox conditions (Figure~\ref{fig:par_converge}Bb,Cb). Estimation of real planets' metallic core fractions enables discussion on their exposure to early collisional processes \citep{charlier_the_2019}, giant impacts \citep{lock_synestia_2018, franco_mercury_2025}, the redox conditions of their deep mantles \citep{lichtenberg_redox_2021, kasting_redox_1993}, and mineralogies \citep{schaefer_ferric_2024, zhang_ferriciron_2024}. Timing arises as a key determiner for accurate insight into exoplanets' potential habitability and deep interiors. The super-solidus conditions suggested by PROTEUS' best-fitting SN solution accurately reproduce the ground-truth's permanent magma ocean regime \citep{wolf_eos_2018, calder_most_2026}. The best-fitting retrieval scenario passes through the ground truth's surface conditions during its evolution (black line in Figure~\ref{fig:static}a), although it eventually overshoots towards cooler surface conditions. The physics resolved by PROTEUS permits ruling against solidified surface conditions, while static InferAGNI retrievals remain agnostic, despite their identical observationally-derived constraints. Planet ages are observationally accessible through stellar dating; e.g., with the asteroseismological component of the upcoming PLATO mission \citep{rauer_plato_2025, aerts_probing_2021}. Our Super-Earth and Terrestrial cases show similar behaviours, so model timing emerges as a key determiner in estimating exoplanet's surface conditions.

The PROTEUS simulations begin simulated evolution from a molten magma ocean state \citep{elkins_linked_2008, tonks_magma_1993} and consider continuous radiogenic internal heat production --- heat from both sources must eventually be radiated to space, through an outgassed atmosphere, for the planets to cool and deflate ---  offset by incoming irradiation \citep{lodders_planetary_1998, noti_effects_2024, nicholls_convective_2025, cmiel_radiative_2025}. Since PROTEUS simulations include a suite of coupled physical processes, they cannot access physically unrealistic conditions that might be permissible for a `free' or reduced-physics retrieval. For example, retaining $\gtrsim 1\mathrm{wt\%}$ low-metallicity atmospheres on small exoplanets for billions of years is likely unrealistic, due to their continuous photoevaporation \citep{owen_evap_2017, rogers_most_2015, moran_high_2023}. Resolving these multiple interacting physical effects means that our prototypical planets may explore different surface conditions to those suggested by static retrieval approaches. Our SN evolutionary retrieval does not permit cool $T_\mathrm{surf}<1250\mathrm{\,K}$ surfaces, while InferAGNI static retrievals do (Figure~\ref{fig:static}a). Physics is a natural `prior' against certain conditions and parameter combinations \citep{macdonald_why_2020}.

In some cases, evolutionary retrievals can be limited by the same degeneracies faced by existing static approaches. This is best demonstrated by our SN case, where the best-fitting scenario's observed radius under-estimates the ground-truth's value (Figure~\ref{fig:obs_converge}Aa). We attribute this behaviour to the planet's extremely reducing interior conditions, which preferentially outgas \ch{H2}-dominated atmospheres. For example, when PROTEUS' ABO samples mantle oxygen fugacities $\Delta \IW=-2$, rather than $\Delta\IW=-3$, the atmospheric radius remains largely unaffected \citep{nicholls_beyond_2026, sossi_redox_2020, arora_thin_2026}. \ch{H2}-preference and decoupling from compositional parameters means that the SN's radius is degenerate with the core fraction and initial volatile budget parameters (Figure~\ref{fig:par_converge}Aa, Ac). Similarly, a given mean molecular weight can be achieved with enhancement of S and C abundances in different proportions (e.g., the SE scenario).

\subsection{Future modelling and development}
\label{sec:discuss_gaps}

Going forward, it will be important to allow models to explore `exotic' parameter regimes. We can identify these theoretically, by searching for unique evolutionary fingerprints, or by using a bootstrapping approach which probes diverse conditions. One option for the latter is to generate large ensembles of models, which sample some prior space, to establish a baseline continuum of exoplanet scenarios \citep{fisher_HowDoWe_2022}. Such an ensemble would then be re-used to bootstrap retrievals of specific exoplanets. Better performance could be derived from the evolutionary retrieval by bootstrapping the ABO algorithm with additional initial samples that probe diverse regimes. 

Our forward-modelling simulates a comprehensive suite of physical and chemical processes, but uses a reduced-complexity configuration of the PROTEUS framework. For example, we use boundary-layer theory to parametrise mantle convection, but future studies should adopt radially-resolved planetary interiors through a mixing-length formalism \citep{bower_numerical_2018, schubert_mantle_2001}. Compositionally fractionating atmospheric escape and orbital evolution might also be considered \citep{herath_thermal_2024, attia_the_2025, bolmont_tidal_2013}. 

`Compute' is a limited resource. Here, explore evolutionary retrievals' efficacy with a limited budget of 100 simulations per retrieval. Section~\ref{sec:results_schemes} found that the performance of ABO retrievals saturates at ten parallel workers. The largest value of the objective function $f$ is not attained sooner when more CPU cores are allocated simultaneously. We adopt a fixed 100-simulation budget per retrieval, so this behaviour is expected: our batch worker-dispatch BO algorithm must select query locations without information of PROTEUS outputs from other batch members. Naturally, the larger the batch size, the more information is missing at query selection, leading to eventually diminishing returns of additional workers \citep{contal2013parallel}. Increasing the total compute budget per retrieval may lead to better fitting solutions, but sensitivity tests with 200 simulations found negligible improvement.  A substantial expansion of the allocated `compute' budget could enable larger numbers of simultaneous workers \textit{without} compromising the number of batches, yielding improved retrieval performance and accuracy.

That CPU-scaling performance saturates before ten simultaneous workers are allocated highlights three opportunities. Firstly, the opportunity for exoplanet retrievals on population scales, where each PROTEUS retrieval instance uses few CPU cores, but multiple instances are applied in parallel to exoplanets from large surveys. The upcoming Roman Space Telescope is expected to identify 100\,000 planets with its wide field imager \citep{johnson_roman_2020}, while PLATO's nominal 4~year lifetime will yield at least 500 Earth-sized planets \citep{matuszewski_plato_2023, rauer_plato_2025}. Secondly, the development of multi-planet retrievals that inform comparative planetology studies. Calculation of our scalar objective $f$ could include observables from multiple planets within a single system (e.g., TRAPPIST-1); the planets' Class\,P2 parameters would be jointly retrieved by simulating their evolution with individual PROTEUS simulation (across multiple CPU cores) collected by a single worker process. Thirdly, an opportunity for accessible retrieval scaling to high-performance computing platforms by each individual PROTEUS simulations utilising multiple CPU cores, reduce each simulation's individual wall-clock runtimes and, correspondingly, reducing the runtime of a complete PROTEUS evolutionary retrieval.

\subsection{Applicability to real systems}
\label{sec:discuss_real}

We present a proof-of-concept test on whether expensive multi-physics planetary evolution simulations can enable planetary lifetime retrieval. Section~\ref{sec:results_bo} demonstrated the feasibility of our approach using three prototypical planets representative of populations within the current exoplanet census (Figure~\ref{fig:exoplanets}; \citet{valencia_diversity_2025, rogers_most_2015}). These prototypes have different masses and compositions, exposed to contrasting irradiation environments, so our demonstration remains agnostic to the unknown nature of real exoplanets. 

Future work should leverage PROTEUS evolutionary retrievals against real exoplanet cases by substituting the parameters and observables adopted here (Table~\ref{tab:var_defs}) with those of real planets, and/or expanding them with different constraints. Our SN prototype can be applied to explain observed divergences between the young sub-Neptunes V\,1298\,Tauri\,c/d, TOI-1136\,b/c, and TOI-451\,b \citep{barragan_mass_2026, beard_the_2024, murphy_kronos1_2026, murphy_kronos2_2026}. Our SE case is applicable for interpreting recent detections of molecular species in the atmospheres of super-Earths L\,98-59\,c/d, TOI\,2076\,e, and Kepler-138\,c/d  \citep{cadieux_l9859_2025, barber_tess_2025, piaulet_evidence_2023, gressier_hints_2024}. Detections of young Earth-sized planets remain limited, but our irradiated Terrestrial scenario is analogous to HD\,63433\,d and Kepler-138\,b \citep{capistrant_hd63433d_2024, jontof_the_2015}. The incorporation of planet age as a constraining quantity on our model may also enable comparative planetology between younger and older systems \citep{beichman_comparative_2007, bean_comparative_2017, murphy_kronos1_2026}.

It is understood that low-mass exoplanets --- particularly super-Earths --- are strongly shaped by their stellar environment through the physics already resolved by PROTEUS \citep{rogers_most_2015, postolec_atmospheric_2026}. Simultaneous planet-star parameter retrieval against combined constraints from stellar and planetary observables may provide additional insights or alleviate ongoing parameter degeneracies \citep{rajpaul_activity_2015, klein_one_2022, martinez_stellar_2017, meech_application_2022}.

\section{Conclusion} 
\label{sec:conclude}

Intimate insights into planets' earliest histories and deep interiors remain inaccessible to static forward models within current retrieval frameworks. Yet, a wealth of incoming data from PLATO, JWST, and the ELT demand interpretation \citep{rauer_plato_2025, gordon_jwst_2022, padovani_elt_2023}. We have tested whether exoplanet lifetimes can be inferred from present-day observations, testing whether parameter retrieval with expensive planetary evolution models is feasible, through the modular PROTEUS framework.

Our main conclusions are summarised below.

\begin{enumerate}
    \item We develop a novel machine learning retrieval framework, founded upon Bayesian optimisation, to asynchronously dispatch computationally-expensive models that resolve planetary evolution across billions of years. Our modelling tool (PROTEUS) includes the interior-atmosphere physics connecting planet's initial conditions to their currently observed states.
    
    \item Three representative exoplanet prototypes test the efficacy of evolutionary retrievals: a young sub-Neptune, an older super-Earth, and a warm Terrestrial. Application of asynchronous Bayesian optimisation to these scenarios suggests that exoplanets' mantle redox states and their early post-formation sulfur, carbon, and hydrogen inventories can be jointly constrained by current spectroscopic observations. 
    
    \item The efficacy of evolutionary retrievals for interpreting oxidised planets' observables may mitigate ongoing observational challenges in reliable measurements of Earth-sized exoplanets' conditions. Oxidised and low-mass regimes are suitable for precise and accurate characterisation, via physics-determined observable-parameter relationships, given sufficient observations.

    \item Comparing observational markers against time-resolved simulations of exoplanets' atmospheric compositions and structures enables real planets' current ages --- through simulation endpoints  --- to be leveraged for accurate parameter estimation. 
    
    \item Evolutionary retrievals are no panacea; accessing sub-Neptunes' internal structures, core fractions, and early volatile budgets remains a challenge. The incorporation of additional physics into forward models alleviates degeneracies faced by static-structure approaches, by being limited to realistic, physically permissible thermal-compositional states.

    \item Machine learning algorithms best suited for this task adopt a Logarithmic Expected Improvement acquisition function, without requiring penalisation-based or Monte-Carlo techniques. Asynchronous Bayesian optimisation for exoplanets' properties can operate within hours of wall-clock time on fewer than 10 CPU cores.

\end{enumerate}

Our demonstration of evolutionary retrievals highlights the utility of process-complete forward modelling. Incorporation of additional physics into PROTEUS may enable improved and accurate understanding of real exoplanets' past conditions. For example, since efficient retrieval is feasible with few CPU cores, jointly modelling secular dynamics of multi-exoplanet systems permits comparative planetology and constraining orbital histories \citep{yu_a_2025, muller_orbital_2018, herath_thermal_2024}. Going forward, leveraging large simulation ensembles to provide the `bootstrapping' initial guesses for retrievals on specific exoplanets may further alleviate parameter degeneracies.

Building upon the current interpretive paradigm, retrievals which adopt simulations of planetary evolution can synergise with the Roman Space Telescope (2026) and the PLATO (2027) mission's combined survey of planet bulk properties, atmospheric compositions, and age-dating. Pairing these upcoming missions with sufficient interpretive tools, resolving the physics that couples planets' deep interiors with their observable atmospheres, will maximise their science return.


\begin{acknowledgments}
H.N. acknowledges support from STFC grant UKRI1184. H.N. thanks Oliver Shorttle for his guidance during this research, and thanks Karen Stuitje for her comments on this manuscript. V.F. was supported by a Branco Weiss fellowship. T.L. was supported by the Branco Weiss Foundation, the Alfred P.\ Sloan Foundation (AEThER project, G202114194), NASA's Nexus for Exoplanet System Science research coordination network (Alien Earths project, 80NSSC21K0593), and the European Research Council (ERC) under the European Union's Horizon Europe research and innovation programme (101219807, MagmaWorlds). R.C. thanks the Science and Technology Facilities Council (STFC) for the PhD studentship (grant reference ST/Y509139/1). We thank the Center for Information Technology of the University of Groningen for their support and for providing access to the Habrok high performance computing cluster.

CRediT author statements. 
\textbf{HN}: Conceptualization, Methodology, Software, Investigation, Data Curation, Writing - Original Draft, Writing - Review \& Editing, Visualization. 
\textbf{TL}: Conceptualization, Resources, Writing - Review \& Editing, Project administration, Funding acquisition. 
\textbf{BR}: Methodology, Software, Writing - Review \& Editing. 
\textbf{RC}: Methodology, Software, Writing - Review \& Editing.
\textbf{VF}: Conceptualization, Supervision, Funding acquisition, Writing - Review \& Editing. 
\end{acknowledgments}

\software{PROTEUS \citep{nicholls_proteusZENODO_2026},  
          AGNI \citep{nicholls_agniZENODO_2026}, 
          SOCRATES \citep{edwards_efficient_1996, manners_fast_2024},
          CALLIOPE \citep{sossi_redox_2020, nicholls_redox_2024},
          MORS \citep{johnstone_active_2021},
          NumPy \citep{harris_array_2020},
          Matplotlib \citep{hunter_matplotlib_2007},
          PyTorch \citep{paszke_pytorch_2019},
          Python,
          Julia \citep{julialang}.
          }

\appendix

\section{Boundary layer mantle dynamics model}
\label{app:boundary}

Our boundary-layer mantle dynamics parametrisation \citep{calder_k218b_2026} solves for these planets' evolving internal potential temperatures $T_\mathrm{pot}$,
\begin{equation}
    \frac{4}{3}\pi c_{p,m} \left(R_\mathrm{int}^3-R_\mathrm{sol}^3\right)\frac{dT_\mathrm{pot}}{dt}=4\pi R_\mathrm{int}^2(F_c-F_l+F_r),
    \label{eqn:tpdiffeqn}
\end{equation}
which are initialised at $T_\mathrm{pot}(t=0)=3200\mathrm{\,K}$ to ensure fully-molten initial states. $c_{p,m}$ is the mantle's specific heat capacity, $\rho_m$ is the density of the mantle, $R_\mathrm{obs}$ is the planet radius, $R_\mathrm{sol}$ is the radius of the solidification front. Heat fluxes are radially-integrated quantities measured at the surface: mantle convective heat transport $F_c$, latent heat release from phase change of crystallisation $F_l$, and radioactive decay $F_r$. The potential temperature $T_\mathrm{{pot}}$ represents an adiabatic mantle temperature structure referenced against the surface conditions. 

We incorporate the latent heat released by the phase change of silicate mantle crystallisation,
\begin{equation}
    F_l = (R_\mathrm{sol}/R_\mathrm{int})^2\Delta H_\mathrm{f}  \rho_m \frac{dR_\mathrm{sol}}{dt},
    \label{eqn:latentheatrelease}
\end{equation}
where $\Delta H_f=\SI{4e6}{\joule\per\kilo\gram}$ is the MgSiO$_3$ specific heat of fusion of  \citep{schaefer_magma_2016}. In parallel, we calculate the radiogenic heat flux $F_r$ assuming that radioisotopes are homogeneously distributed throughout the mantle at Earth-like abundances \citep{lodders_planetary_1998,schaefer_water_2015}. 

The mantle convective heat flux $F_c$ is calculated using a boundary-layer theory parametrisation of Rayleigh-Bernard convection, in the `soft turbulence' regime \citep{lebrun_thermal_2013, hamano_lifetime_2015,schaefer_magma_2016, Meier2023, tackley_tectono_2023}. Numerical and empirical experiments have determined that the convective heat flux $F_c$ scales as a function of the Rayleigh number and temperature contrast \citep{schubert_treatise_2015},
\begin{equation}
    F_c = \frac{k_c}{l}(T_\mathrm{pot}-T_\mathrm{surf})\Big(\mathrm{\frac{Ra}{Ra_{cr}}}\Big)^n.
\end{equation}
The fluid undergoes laminar flow at small Ra$~\sim10^2$ and turbulent flow at large Ra$\sim10^{30}$. We adopt an exponent $n=1/3$ and thermal conductivity $k_c=\SI{4.2}{\watt\per\meter\per\kelvin}$ from empirical measurements \citep{turcotte_geodynamics_2002}. The length-scale $l$ is set equal to the interior (mantle) radius $\mathrm{R_{int}}$. We use a value of 1100 for the critical Rayleigh number \citep{schaefer_magma_2016}.

The Rayleigh number, Ra, characterises the relative importance of convective energy transport compared to thermal diffusion \citep{turcotte_geodynamics_2002}.
\begin{equation}
    Ra=\frac{\rho_m \alpha g\left(T_\mathrm{pot}-T_{\mathrm{surf}}\right)l^3}{\kappa \eta},
    \label{eqn:rayleighnumber}
\end{equation}
where $T_\mathrm{pot}$ is the mantle potential temperature, $\alpha$ is the thermal expansibility of the mantle, $g$ is the surface gravity, $\kappa$ is the thermal diffusivity and $\eta$ is the dynamic viscosity. 

Following \citet{lebrun_thermal_2013} and \citet{schaefer_magma_2016}, we use the Vogel-Fulcher-Tammann relation to analytically represent the mantle's dynamic viscosity as a function of the potential temperature:
\begin{equation}
    \eta = \eta_0 \frac{\exp [E_a/(R_\mathrm{gas} T_\mathrm{pot})]}{(1-(1-\Phi_v)/(1-\Phi_v^\mathrm{crit}))^{5/2}},
\end{equation}
where the denominator deviates from an Arrhenius-like activation function to adjust for the entrained crystal fraction \citep{ikeda_understandi_2013}. The dynamic viscosity is normalised to \SI{3.8e9}{\pascal\second} in the solid-phase $\Phi_v=0$ end member \citep{schaefer_magma_2016}.

The mantle's volume melt fraction is calculated from its potential temperature,
\begin{equation}
    \Phi_v=\frac{T_\mathrm{pot}-T_{\mathrm{sol}}}{T_{\mathrm{liq}}-T_{\mathrm{sol}}},
    \label{eqn:meltfrac}
\end{equation}
where $T_{\mathrm{sol}}=1420\mathrm{\,K}$ and $T_{\mathrm{liq}}=2020\mathrm{\,K}$ are the solidus and liquidus temperatures \citep{lebrun_thermal_2013}. 

We assume that the magma ocean undergoes fractional bottom-up monotonic solidification \citep{solomatov_nonfractional_1993, bower_linking_2019, maurice_onset_2017, monteux_cooling_2016}, which allows expressing the solidification front's radius as a function of the volumetric melt fraction, \begin{equation}
    R_\mathrm{sol}^3=R_\mathrm{int}^3-\Phi_v \cdot (R_\mathrm{int}^3-R_\mathrm{core}^3),
    \label{eqn:radiussolidification}
\end{equation} 
and an analytic expression for its time derivative,
\begin{equation}
    \frac{d R_\mathrm{sol}}{dt}=-\frac{R_\mathrm{int}^3-R_\mathrm{core}^3}{3R_\mathrm{sol}^2\left(T_{\mathrm{liq}}-T_{\mathrm{sol}}\right)}\frac{dT_\mathrm{pot}}{dt}.
    \label{eqn:drsdt}
\end{equation}

The surface temperature $T_\mathrm{surf}$ evolves subject to balance between the energy transported through the atmosphere ($F_{\mathrm{atm}}$, calculated by AGNI) and the energy arising from the planet's interior,
\begin{equation}
\begin{split}
    \left[c_{p,\mathrm{atm}}M_{\mathrm{atm}}+c_{p,m}\frac{4}{3}\pi\rho_m\left(R_\mathrm{int}^3- \left(R_\mathrm{int}-\delta\right)^3\right)\right]\frac{dT_{\mathrm{surf}}}{dt}= \\ 4\pi R_\mathrm{int}^2\left(F_c-F_{\mathrm{atm}}\right),
    \label{eqn:tsurfdiffeqn}
\end{split}
\end{equation} 
where $c_{p,\mathrm{atm}}$ is specific heat capacity of the atmosphere, $M_{\mathrm{atm}}$ is the mass of the atmosphere (calculated by PROTEUS' outgassing module, CALLIOPE), $\delta$ is the thickness of the surface conductive boundary layer and  is the net flux from the atmosphere (calculated by PROTEUS' atmosphere module, AGNI). The thickness of the thin solidified boundary layer at the surface is set by conduction,
\begin{equation}
    \delta=k\left(T_\mathrm{pot}-T_\mathrm{surf}\right)/F_c
    \label{eqn:boundarylayerthickness}
\end{equation}

Core radius fractions $r_c=R_\mathrm{core}/R_\mathrm{int}$ are held constant in time --- although they are varied between the explored planet scenarios --- and solve for their average core densities using analytical scaling relationships \citep{noack_parameteris_2020}. The core material's specific heat capacity is held constant at \SI{880}{\joule\per\kilo\gram\per\kelvin} \citep{lodders_planetary_1998, bower_numerical_2018}.

\section{Asynchronous Bayesian optimisation algorithm}
\label{app:algo}

Here, we formally describe the algorithmic procedure by which PROTEUS implements batched asynchronous Bayesian optimisation (ABO). Algorithm~\ref{alg:asyncBO} uses $q$ workers, each allocated a single CPU core, which each supports one PROTEUS simulation  (Section~\ref{sec:methods_proteus}) to perform a query (a forward model) in parameter space.

\begin{algorithm}[!ht]
    
    \begin{algorithmic}[1]
    
    \REQUIRE Oracle $f(\cdot)$, acquisition function $\alpha(\cdot)$, globally stored initial data $D_0 = \{(x_i, y_i)\}_{i=1}^{n_0}$, query budget $N$
    \STATE \textbf{initialise:} $\mathcal{B} \leftarrow $ $\{x_{j}\}_{j=1}^q$ quasi-random
    \STATE \textbf{spawn:} $q$ worker processes
    \FOR{$x_j \in \mathcal{B}$ }
    \STATE start worker $j$ with query $f(x_j)$
    \ENDFOR
    \STATE \textbf{set:}  $n \leftarrow 0$
    \WHILE{$n < N$}
    \IF{ worker $j$ completes query $(x_{n+1}, y_{n+1})$} 
    \STATE $D_{n+1} \leftarrow D_n \cup \{(x_{n+1}, y_{n+1})\}$
    \STATE $\mathcal{B} \leftarrow \mathcal{B} \setminus \{x_{n+1}\}$
    \STATE $\mathcal{GP}_j \leftarrow$ fit-surrogate($\mathcal{GP}_j$, $D_{n+1}$)
    \STATE $x' \leftarrow \arg\max \limits_{x \in \mathcal{X}} \ \alpha(x \mid D_{n+1})$ 
    \STATE $\mathcal{B} \leftarrow \mathcal{B} \cup \{x'\}$
    \STATE start worker $j$ with query $f(x')$
    \STATE $n \leftarrow n +1$
    \ENDIF
    \ENDWHILE
    \RETURN $i^* = \arg \max_{i} y_i$ and $(x_{i^*}, y_{i^*})$
    \end{algorithmic}

    \caption{Asynchronous BO with $q$ workers, implemented within the PROTEUS framework.}
    \label{alg:asyncBO}
    
\end{algorithm}

\section{Thermal and compositional evolution of ground-truth scenarios}
\label{app:truthevo}

\begin{figure*}
    \centering
    \includegraphics[height=0.350\linewidth]{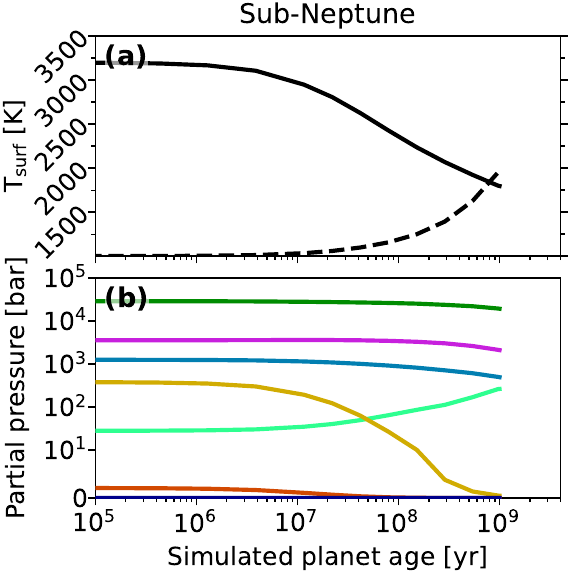}%
    \includegraphics[height=0.345\linewidth]{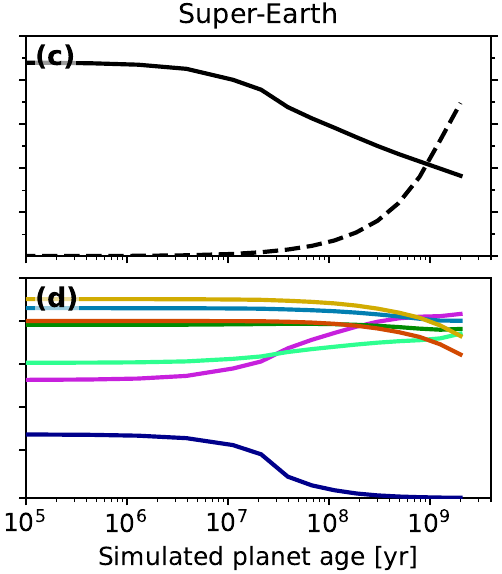}%
    \includegraphics[height=0.350\linewidth]{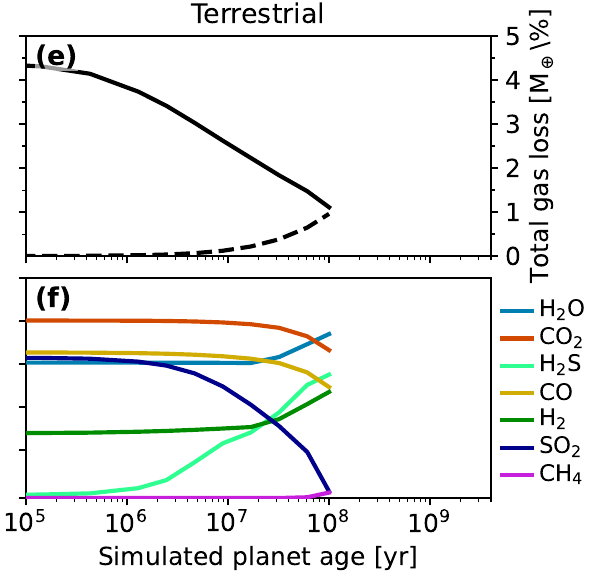}%
    
    \caption{Thermal and compositional evolution of ground-truth scenarios: Sub-Neptune (\textbf{left}), Super-Earth (\textbf{middle}), and Terrestrial (\textbf{right}). \textbf{Top}: surface temperature (solid) and the time-integrated amount of gas lost (dashed) as the planets evolve from their initial state (x-axis). \textbf{Bottom}: partial surface pressures of major gas species (line colours), on the same time-axis as top panels.}
    \label{fig:truthevo}
\end{figure*}

Prototype exoplanet cases simulated with PROTEUS (Section~\ref{sec:methods_proteus}) establish the ground-truth observables against  which we test the efficacy of retrieving exoplanets' parameters using PROTEUS. Figure~\ref{fig:truthevo} shows the thermal (top) and compositional (bottom) evolution of our three prototypes (columns). These planets differ by their Class\,P1~and~P2 input parameters, which yield their different Class\,O1~and~O2 observables (Table~\ref{tab:var_defs}).

All three scenarios cool from an initially hot state, in which their mantles are fully molten. Their surface temperatures exceed 3000\,K at this point (Figure~\ref{fig:truthevo} top panels, solid lines). All rocky planets are expected to begin with primordial magma oceans, through the energy obtained during their accretion \citep{elkins_review_2012, halliday_review_2023}, formation \citep{deeg_formation_2018, lichtenberg_geophysical_2022}, and metallic core segregation \citep{wade_core_2005, karato_core_1997}. The surface temperatures $T_\mathrm{surf}$ decrease monotonically as the planets thermally radiate energy to space. Emission from their surfaces is modulated by atmospheric blanketing, since their surface pressures ($>100$\,bar) generate optically-thick radiation streams \citep{nicholls_redox_2024, abe_early_1986}. The atmospheric temperature structures and gas opacities also induce a strong greenhouse effect \citep{kasting_runaway_1988, pierrehumbert_greenhouse_2011, hamano_emergence_2013, kopparapu_habitable_2013}. Outgoing thermal emission is offset by incoming stellar radiation \citep{johnstone_active_2021, spada_radius_2013}, which is modulated by an upward radiation stream from Rayleigh scattering \citep{pierrehumbert_book_2010}.

The modelled atmospheric partial pressures evolve due to multiple simultaneously-acting processes. The bottom row of Figure~\ref{fig:truthevo} plots the volume mixing ratios of near-surface gas phase, for each case (columns). These evolve through the gradual stripping of CHNS elements by hydrodynamic escape, temperature dependent shifts in thermochemical speciation, and repartitioning between the atmosphere and magma ocean. 

\section{Best-fitting estimates of parameter and observable quantities}
\label{app:bestfit}

Our asynchronous Bayesian optimisation algorithm is configured to perform 100 PROTEUS simulations per retrieval, to estimate the best-fitting values of our Class\,P2 parameters, given some constraining Class\,O1 observables. Each PROTEUS simulation also calculates a large number of Class\,O2 output quantities --- from which we highlight the atmospheric \ch{H2O} volume mixing ratio, surface temperature, and surface pressure. These values underpin the results presented in Sections~\ref{sec:results_bo}~and~\ref{sec:results_static}.

This appendix section quantifies the parameters, observables, and several of the output quantities corresponding to the best-fitting scenarios explored by PROTEUS' ABO retrieval algorithm. Table~\ref{tab:best_fit} presents the best-fitting scenarios' estimated Class\,P2 parameters, constraining Class\,O1 observables, and output quantities. Errors are quantified linearly against the synthetic ground-truth (Table~\ref{tab:var_defs}) --- we emphasise that several quantities are physically expected to vary across multiple orders of magnitude, although errors are quantified on a linear scale.

\begin{table*}[h]
\centering
\begin{tabularx}{\linewidth}{lllll}
    \hline
    \textbf{(P2)} Estimated parameter               & Sub-Neptune              & Super-Earth             & Terrestrial          & Expected scaling behaviour \\ 
    \hline
    Metallic core frac. {[}\% radius{]}    & 46.9 (25.4\%)  & 30.0 (25.0\%) & 63.0 (12.7\%) & Linear       \\
    $f$O$_2$ {[}$\Delta\IW${]}             & -1.44 (52.0\%) & +1.19 (58.1\%) & +4.00 (25.0\%) & Linear       \\
    Initial $\Hppmw$                       & 2950 (70.5\%)  & 1700 (83.0\%) & 1570 (21.3\%) & Logarithmic          \\
    Initial C/H                            & 1.60 (68.8\%)  & 1.30 (13.7\%) & 2.72 (8.2\%)  & Logarithmic          \\
    Initial S/H                            & 0.86 (6.7\%)   & 3.17 (37.0\%) & 1.01 (21.2\%) & Logarithmic          \\ 
    \hline
    \textbf{(O1)} Constraining observable                        & Sub-Neptune              & Super-Earth             & Terrestrial          & Expected scaling behaviour \\ 
    \hline
    Photosphere radius {[}$R_\oplus${]}    & 1.52 (21.1\%)  & 1.34 (8.3\%)  & 1.03 (4.8\%)  & Logarithmic          \\
    Photosphere temp. {[}K{]}              & 314 (0.2\%)   & 365 (10.7\%) & 576 (1.1\%)  & Linear       \\
    Photosphere gravity {[}m/s$^2${]}      & 1.29 (37.3\%)  & 1.05 (13.6\%) & 9.44 (9.1\%)  & Logarithmic          \\
    Atmosphere molec. weight {[}g/mol{]}   & 10.3 (60.4\%)  & 18.0 (4.8\%)  & 25.9 (0.6\%)  & Linear       \\
    Atmosphere C/O (wt.)                   & 5.21 (38.7\%)  & 0.02 (97.1\%) & 0.17 (11.1\%) & Logarithmic          \\
    Atmosphere S/O (wt.)                   & 1.00 (5.9\%)   & 0.61 (6.7\%)  & 0.10 (14.8\%) & Logarithmic          \\
    Atmosphere O/H (wt.)                   & 0.35 (54.1\%)  & 4.95 (43.4\%) & 13.20 (1.2\%)  & Logarithmic          \\ 
    \hline
    \textbf{(O2)} Extra output quantity                   & Sub-Neptune              & Super-Earth             & Terrestrial          & Expected scaling behaviour \\ 
    \hline
    \ch{H2O} volume mix. ratio {[}\%{]} & 6.3 (64.3\%)  & 58.4 (59.1\%) & 65.9 (7.6\%)  & Logarithmic          \\
    $P_\mathrm{surf}$ {[}kbar{]}           & 1.20 (46.2\%)  & 1.95 (95.5\%) & 4.82 (41.4\%) & Logarithmic          \\
    $T_\mathrm{surf}$ {[}K{]}              & 1630 (9.5\%)   & 1350 (29.6\%) & 1380 (11.4\%) & Linear            
\end{tabularx}
\caption{Best-fitting parameters, observables, and additional output variables estimated by PROTEUS' ABO retrievals for our three prototype exoplanet cases. Physics dictates that some variables scale across logarithmically, across multiple orders of magnitude, while others will exhibit linear behaviours. We quantify the relative misfit `error' compared to the ground-truth (Table~\ref{tab:var_defs}) in parentheses: $\epsilon=|(o-t)/t|$.}
\label{tab:best_fit}
\end{table*}

\section{Forward-model computational expense}
\label{app:evaltime}

Our machine learning retrieval mode (within the PROTEUS framework) enables the application of a computationally expensive simulations (forward models) through asynchronous Bayesian optimisation \citep{riegler2026standard}. The runtime of each PROTEUS simulation is variable. We adopt an adaptive time-stepping scheme to run simulations up to a targeted integration time, but simulations may terminate before this point (Section~\ref{sec:methods_proteus}) and numerical performance depends on the `stiffness' of the scenario modelled \citep{hindmarsh_scientific_1983, press_numerical_2007, julialang}. Each PROTEUS simulation was run on a single CPU thread, but multiple workers and retrieval instances enable parallelisation. We used a Intel~Xeon~E5-2650~v4 CPU with 48~cores that have a maximum frequency of 2.9~GHz.

\begin{figure}
    \centering
    \includegraphics[width=0.5\linewidth]{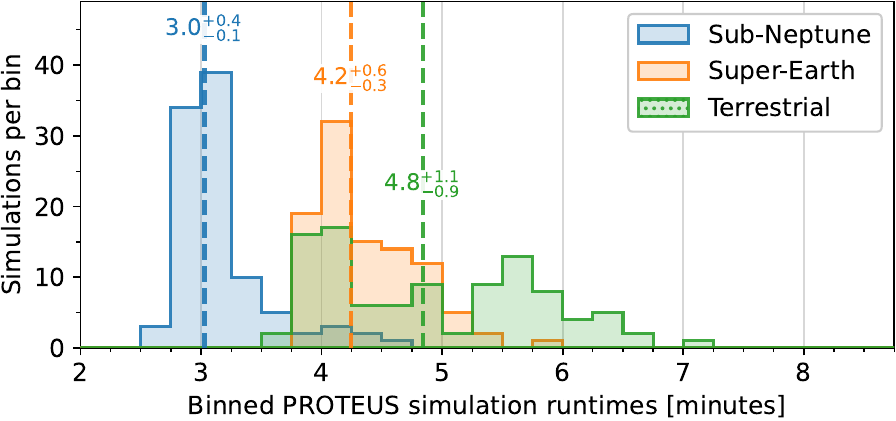}
    \caption{Histograms of PROTEUS simulation wall-clock runtimes (minutes), corresponding to the baseline asynchronous BO retrievals presented in Figures~\ref{fig:obs_converge}~and~\ref{fig:par_converge}. Median values are shown by dashed vertical lines; annotations include median and $\pm1$ standard deviation ranges. Histogram colours denote the exoplanet case being modelled. A total of 100 simulations were run for each retrieval.}
    \label{fig:evaltime}
\end{figure}

Figure~\ref{fig:evaltime} plots histograms of PROTEUS wall-clock runtimes seen during our baseline retrievals of the Sub-Neptune, Super-Earth, and Terrestrial scenarios (histogram colour; Section~\ref{sec:results_bo}). PROTEUS wall-clock runtimes range from 2~to~8 minutes, but depend strongly on the targeted physical regime. The TR simulations typically had the shortest wall-clock runtime ($\sim3$~minutes), consistent with it having the shortest integration time. The SE simulations (orange) had the longest maximum integration time, but their wall-clock runtimes are generally shorter than the SN case (blue). Simulations within the sub-Neptune regime are computationally more expensive, because radiative-convective solutions for their \ch{H2}-dominated atmospheres require more iterations to converge \citep{nicholls_agni_2025}.

The incorporation of additional physics into the forward-model simulations would increase these wall-clock runtimes. For example, by using the SPIDER interior geodynamics model rather than a boundary-layer parametrisation, which would provide enhanced model accuracy in the mixed-phase magma ocean regime \citep{bower_linking_2019, abe_thermal_1993}.

\bibliography{main}{}
\bibliographystyle{aasjournal}

\end{document}